**Uncooled Carbon Nanotube Photodetectors**

*Xiaowei He, François Léonard\*, and Junichiro Kono\**

X. He, Prof. J. Kono
Department of Electrical and Computer Engineering
Rice University
Houston, Texas, USA
E-mail: kono@rice.edu

Dr. F. Léonard
Sandia National Laboratories
Livermore, California, USA
fleonar@sandia.gov

Photodetectors play key roles in many applications such as remote sensing, night vision, reconnaissance, medical imaging, thermal imaging, and chemical detection. Several properties such as performance, reliability, ease of integration, cost, weight, and form factor are all important in determining the attributes of photodetectors for particular applications. While a number of materials have been used over the past several decades to address photodetection needs across the electromagnetic spectrum, the advent of nanomaterials opens new possibilities for photodetectors. In particular, carbon-based nanomaterials such as carbon nanotubes (CNTs) and graphene possess unique properties that have recently been explored for photodetectors. Here, we review the status of the field, presenting a broad coverage of the different types of photodetectors that have been realized with CNTs, placing particular emphasis on the types of mechanisms that govern their operation. We present a comparative summary of the main performance metrics for such detectors, and an outlook for performance improvements.

# 1. Introduction

Photodetectors play important roles in many areas of modern life. For example, the size reduction and performance improvements of charge-coupled devices[1] based on silicon has led to the ubiquitous presence of color cameras in portable smart phones, and enabled the era of digital cinematography. Photodetectors that operate in the infrared (IR) spectrum[2] are being used extensively in national security applications such as night vision, remote sensing[3], environmental monitoring, and surveillance. The extension of photodetector technology to the terahertz (THz)[4] has opened new fields, and shows promise in a number of emerging areas such as medical imaging[5], food inspection[6], and security[7].

These many applications of photodetectors and their increasing performance have been enabled by the continuous discovery and improvement of materials. For example, ultra-sensitive midwave-IR detectors are often based on HgCdTe, a material that was first discovered in the 1960s, and soon identified as promising for IR applications due to its small, tunable optical bandgap. More than 50 years later, after many generations of continued improvement, this material is still extensively used. More recently, the improved control over quantum dot and quantum well growth, particularly in III-V materials such as GaAs and InAs, has led to new paradigms for IR detectors. While these technologies have enabled exceptional applications and ever-increasing performance, the exotic nature of the materials can lead to issues of reliability, integration, size, cost, as well as geopolitical and environmental concerns due to the geographical location of these material sources or growth capabilities. This, coupled with the constant search for new and more complex applications, has spurred a tremendous amount of interest in exploring new materials and detection modalities. In particular, nanomaterials, nanodevices, and nanosystems possess unique intrinsic optoelectronic, geometrical, and structural properties that could enable photodetectors with new form factors, tunability, flexibility, reconfigurability, etc.

In addition to nanostructures such as quantum dots and quantum wells, carbon nanotubes (CNTs) and graphene have shown unique promise for a broad range of photodetectors from the ultraviolet to the THz. A recent review[8] of graphene photodetectors has been presented, and we therefore focus this review on CNT photodetectors. We introduce the basic properties of CNTs relevant to optoelectronics and discuss the different types of photodetectors that have been realized. The current performance status of these detectors is evaluated, and their potential for further improvement is discussed.

**1.1. Properties of carbon nanotubes relevant to photodetectors**

Our goal in this section is to establish the basic properties of CNTs that are relevant to photodetectors without going into too many intricate details. Other monographs have extensively discussed CNT properties[9, 10], and we refer the reader to these publications.

*Graphene*. Since the properties of CNTs derive from those of graphene, we begin by discussing the electronic and optical properties of graphene. Structurally, graphene consists of a single layer of carbon atoms arranged in a honeycomb lattice, where each carbon atom makes an $sp^2$-bond with three neighbors. It turns out that the electronic structure of graphene can be understood at the simplest level by using a tight-binding representation for these bonds. The bandstructure near the Fermi level consists of linear crossing bands; however, unlike a conventional metal, the density of states is zero at the Fermi level due to the reduced dimensionality.

The optical absorption in graphene is dominated by different phenomena depending on the wavelength[11], as illustrated in **Figure 1**. In the visible, near-, and mid-IR, interband transitions dominate the absorption, leading to an optical absorption that is independent of wavelength and equal to 2.3%. In the far-infrared and THz, intraband transitions dominate the absorption, which is controllable by electrical gating.[12, 13] It is also possible to harness plasmons in doped graphene for optical absorption, but because the plasmons have finite energy at small momentum it is necessary to use a grating to couple the light with the graphene plasmons.[12, 14] Finally, in the ultraviolet, an absorption peak occurs due to the saddle-point singularity at the M-point, modified by excitonic effects, which typically attains ~10 percent.[11, 15] It is important to note that different absorption mechanisms can occur simultaneously, and that electrostatic gating can be used to tune the carrier concentration and the optical absorption.

Other graphene properties relevant to photodetectors include carrier mobility and carrier recombination times. The carrier mobility in exfoliated and suspended graphene has been shown to be as high as 200,000 cm$^2$/Vs[16] with values around 10,000 cm$^2$/Vs for graphene on substrates. The carrier recombination times depend on both the doping and the presence of impurities, but typical values have been measured around 1-10 ps.

In the context of bolometers, the temperature coefficient of resistance (TCR) is an important quantity, defined as

$$TCR = \frac{1}{R}\frac{dR}{dT}. \tag{1}$$

In graphene, the intrinsic TCR is relatively low, with values of only a fraction of %/K,[17] much lower than typical materials used in existing bolometers.

As will be discussed below, for photothermoelectric detectors, the Seebeck coefficient and thermal conductivity enter the figure of merit. The Seebeck coefficient for graphene depends on doping and temperature: the largest values are reported at room temperature[18], while the dependence on doping follows that expected for ambipolar materials with a value of zero for zero doping, an increase to near 100 µV/K at low doping (room temperature) and a decrease with increasing doping. The thermal conductivity is generally high, with values at room temperature ranging from 600 W/mK for supported graphene to 2500 W/mK for suspended samples.[19]

*Carbon nanotubes*. The optical properties of CNTs, especially those of single-wall CNTs (SWCNTs), can be understood from those of graphene since the atomic structure of CNTs can be obtained by imagining a strip of graphene that is rolled to make a closed cylinder.[9] The quantization of the electron wavefunction around the CNT circumference leads to an electronic bandstructure that depends sensitively on the CNT diameter, and its so-called chirality. At a high level, this implies that CNTs can be semiconducting or metallic, with the semiconductor bandgap depending inversely on the diameter. This, coupled with the one-dimensional nature of CNTs, leads to important differences for the optical properties compared to graphene.

The first important difference is due to the fact that each subband in a CNT has a specific angular momentum, and as a consequence, there are selection rules for optical transitions. For example, optical absorption can be maximized or suppressed depending on the direction of light polarization with respect to the CNT axis. A second important difference arises because of the enhanced Coulomb interaction in quasi-one-dimensional materials. This leads to a significant binding energy for excitons, which dominate the visible, near-infrared, and mid-infrared optical absorption in CNTs. This has important consequences for photodetectors because large enough electric fields must be present to dissociate the exciton in photoconductive or photovoltaic devices. Note that doping of the CNT can impact the excitonic properties,[20, 21] and that it can lead to THz absorption due to plasmons.[22] In the case of metallic CNTs,[23] the interband transitions between the linear bands are not allowed even though the density of states is constant in a broad range of energies around the Fermi level. As a consequence, the optical absorption in the near-infrared, visible, and UV ranges are dominated by excitonically modified interband transitions between parabolic bands, similar to semiconducting nanotubes.

The mobility of semiconducting CNTs depends significantly on the method used to prepare the CNTs and devices. For example, individual CVD-grown CNTs have been shown to have mobilities as large as 10,000 cm$^2$/Vs, while solution-processed CNTs have much lower mobilities that can attain 300 cm$^2$/Vs.[24]

The Seebeck coefficient of metallic and semiconducting CNTs behave similarly to the case of graphene. At zero doping, both semiconducting and metallic CNTs have zero Seebeck coefficient; as the doping increases, the metallic CNTs reach a maximum Seebeck coefficient of 40 μV/K while the maximum for semiconducting CNTs is 300 μV/K.[25] The thermal conductivity of individual CNTs depends on the diameter, temperature, and quality of the CNTs, but typically has values between 1000 and 3000 W/mK.[26]

The above discussion describes the properties of individual CNTs, but for applications, thin films of CNTs are most likely to be adopted. While in some cases the properties of thin films can be considered as those of the individual components, most often this is not the case. For example, most CNT thin films are composed of randomly oriented individual CNTs such that electronic transport is through CNT-CNT junctions. This affects the mobility, the Seebeck coefficient, and the thermal conductivity. However, for disordered thin films, the optical absorption can be considered as a superposition of the individual absorptions. In the case of thin films with aligned and dense CNTs, new effects come into play. For example, collective antenna effects are found in the THz,[27] and CNT-CNT interactions can lead to small shifts of the excitonic transitions.

Carrier recombination processes in CNTs are affected by a numbr of factors, including metallicity, individuality, doping, temperature, and carrier density. In CNT films consisting of mixed semiconeducting and metallic CNTs, the recombination time is less than 1 ps due to the rapid non-radiative decay through metallic CNTs[28]. In individually suspended semiconducting CNTs under weak optical excitations, the carrier recombination time is larger by one order of magnitude,[21] which allows the diffusion length of excitions along CNT channle to be as long as hundreds of nanometers.[29, 30] Under strong optical pumping, however, rapid Auger recombination (exciton-exciton annihilation) occurs,[31] making recombination times shorter with increasing carrier density, especially at high temperatures where exciton diffusion induced collisions are accelarated.[32]

The TCR of CNT films is normally far less than one %/K at room temperature. However, it can be enhanced by type separtion and the incorpation of certain types of polymers. For example, the TCR of the semiconducting-SWCNT/PVP (polyvinylpyrrolidone) composite can be up to ~6.4 %/K,[33] which is higher than the TCR (> 3%/K) of vanadium oxide— the current leading platform for uncooled bolometric detection.[34]

**1.2. Performance metrics for photodetectors**

In this review, we focus on the pixel-level performance of photodetectors as opposed to the image-level performance, and we are thus mainly concerned with the basic efficiency of optical to electrical energy conversion. It is important to re-iterate that quantitative performance metrics must be taken in the context of the maturity of the technology, and other attributes such as ease of integration, cost, power requirements, etc.

The simplest performance metric is the responsivity defined as

$$R_V = \frac{voltage\ generated}{incident\ optical\ power} \quad (2)$$

or

$$R_I = \frac{current\ generated}{incident\ optical\ power} \quad (3)$$

In our case, we take these two measures to be equivalent in the sense that the internal resistance of the device relates the voltage and the current. (We note that sometimes the absorbed power is used in these definitions, but this artificially increases the performance; we do not follow this practice here.)

Other metrics for photodetectors are the external and internal quantum efficiencies,[35] defined as,

$$EQE = \frac{photocurrent\ /\ charge\ of\ one\ electron}{incident\ optical\ power\ /\ energy\ of\ one\ photon}, \quad (4)$$

$$IQE = \frac{EQE}{1 - reflection - transmission}. \quad (5)$$

Besides the Responsivity, *EQE*, and *IQE*, additional performance parameters for photodetectors working at zero bias also include the short-circuit current, $I_{sc}$, open cicuit voltage, $V_{oc}$, fill factor (*FF*), and power conversion efficiency ($\eta$).[36] The fill factor and power conversion efficiency are defined by

$$FF = \frac{I_M V_M}{I_{SC} V_{OC}} \quad (6)$$

$$\eta = \frac{I_M V_M}{P_{in}} \quad (7)$$

where $I_M$ and $V_M$ describe the values of photocurrent and photovoltage at which the output power reaches its maximum, and $P_{in}$ is the incident power density.

While the above parameters give insight into the efficiency of the photoresponse mechanisms, for photodetectors what matters is the signal-to-noise ratio. Thus, a more complete metric for photodetectors is the noise-equivalent-power (NEP) defined as

$$NEP = \frac{noise\ spectrum}{responsivity} \qquad (8)$$

where the noise spectrum has units of $V/\sqrt{Hz}$ for a responsivity $R_V$, so that the *NEP* has units of $W/\sqrt{Hz}$. This value defines the smallest power that can be detected over the noise. An equivalent metric that normalizes the *NEP* by the active area of the detector $A$ is the specific detectivity

$$D^* = \frac{\sqrt{A}}{NEP}. \qquad (9)$$

(Note that this definition goes together with the above definition of *NEP*; a common alternative approach is to include the detector bandwidth in the definition of $D^*$, in which case the definition and units of *NEP* are different.) Finally, the speed of response is also important when selecting a detector for a particular application. Here we define the speed of response from either the rise or fall time of the photosignal $\tau$, or the cut-off frequency $f_c$; the two are related as $f_c^{-1} = 2\pi\tau$.

**1.3. Types of photodetectors**

We classify photodetectors into five types based on the fundamental photoresponse mechanism – photoconductive, photovoltaic, photogating, bolometric, and photothermoelectric, corresponding to photodetectors known as photoconductors, photodiodes, phototransistors, bolometers, and photothermopiles, respectively.

*Photoconductive*: In *photoconductors*, optically-created electron-hole pairs are separated in the channel due to electric fields created by an applied source-drain bias. The I-V curve under illumination has the same shape as the one in the dark, but multiplied by a constant factor. In particular, there is no photocurrent generated at zero bias, as illustrated in **Fig. 2**.

*Photovoltaic*: The photovoltaic effect that occurs in *photodiodes* is similar to the photoconductive case, except for the presence of a net electrical field in the absence of a source-drain bias. As a consequence, a photocurrent is observed at zero bias. This mechanism is usually observed in rectifying devices, which can be realized with a built-in p-n junction or with an asymmetric-contact device where one of the contacts is of Schottky type while the other is ohmic. (A photovoltaic effect can also be observed in back-to-back Schottky contacts if only one of the contacts is illuminated.) The photovoltaic effect is also the mechanism found in solar cells.

*Photogating*: In a typical field-effect transistor configuration, a gate voltage is used to modulate the current flow between source and drain electrodes, with a sharp transition between the ON and OFF states of the transistor. Any optically-induced change in the effective gate voltage will cause a change in the source-drain current that can be electrically detected. This type of photodetector is called a *phototransistor*. In this case, one is typically interested in the change of the transfer characteristics of the FET upon illumination, which is usually manifested as a shift of the transfer curve.

*Bolometric*: In *bolometers*, the absorbed light increases the temperature of the active material causing a change in its resistance. The resistance can increase or decrease depending on the temperature coefficient of resistance (TCR) of the active material. The I-V curves do not show a photocurrent at zero bias, and the curves look similar to the photoconductive case if the TCR is negative.

*Photothermoelectric*: This effect is related to the bolometric effect in the sense that the impact of the light absorption is to increase the temperature. However, in this case, a photocurrent is obtained by exploiting the Seebeck effect where the application of a temperature gradient across a thermoelectric material generates a voltage. This type of photodetector is called *photothermopiles* (or sometimes just *thermopiles*). The different ways in which the photothermoelectric effect can be harnessed for photodetection will be discussed further below, but a general signature is an I-V curve that is rigidly shifted upon light absorption. This is similar to the photovoltaic effect in that a photocurrent is generated at zero bias, except that no rectifying I-V curve is required.

## 2. CNT Phototransistors

The area of CNT phototransistors is the least mature in terms of devices that are relevant for applications. CNT transistors were used early on to study the basic photophysics of individual CNTs, but few studies have actually demonstrated a phototransistor action. By this we mean a mechanism where the threshold voltage of the transistor is shifted due to illumination, as discussed in **Fig. 2**.

**Table 1** Summary of CNT phototransistors.

| Materials | Arctechture | Response time | Responsivity | Detectivity [cm Hz$^{1/2}$/W] | Wavelength [um] |
|---|---|---|---|---|---|
| Individual s-SWCNT | FET | 5 s | $10^4$ A/W | N/A | 0.3-0.6 |
| SWCNT/C$_{60}$ thin film | FET | 2-4 ms | 97.5 A/W | $10^9$ | 1-1.4 |

A few examples have appeared where charge traps in the underlying substrate can lead to photogating[37], but such phenomena are difficult to control and implement in real applications. More promising are approaches that intentionally introduce photosensitive elements in close proximity to the CNTs that serve as effective gates. One example of this approach is to functionalize the CNTs with molecules such as azobenzenes that undergo a cis-trans transition upon light absorption, accompanied with a large change in dipole moment [38] which serves as an additional gate for the CNT transistor. This approach has demonstrated a very high responsivity of $10^4$ A/W, with spectral selectivity using appropriately designed chromophores, but so far it has only been applied to individual CNT devices in the visible range. A related approach was recently demonstrated using layers of C$_{60}$ deposited on CNT thin film transistors[39] with a detectivity of $10^9$ cm Hz$^{1/2}$/W in the near-IR. Quantum dots [40] with absorption in the visible have also been used in the same spirit.

## 3. CNT Photoconductors

A major problem for SWCNT photoconductors is that photo-induced electron-hole pairs have strong binding energies on the order of several hundred of meV,[41] which exceeds the room temperature thermal energy. Consequently, the separation of electron hole-pairs is inefficient without an strong electric field, and therefore curtails the detector performance.

**Table 2** Summary of major CNT photoconductive detectors from 2003 to 2014[a]

| TMaterials | Architecture | EQE (IQE) | Response Time | Responsivity | Dectivity [cm Hz$^{1/2}$/W] | Wavelength [μm] |
|---|---|---|---|---|---|---|
| Individual s-SWCNT[42] | FET | ~10% | N/A | ~0.1 pA/W | N/A | 0.78-0.98 |
| SWCNT/PC[43] | Composite | N/A | 40 ms | ~0.4 μA/W | N/A | 1.3 |
| s-SWCNTs /P3HT[44] | BHJ[b] | (86%) | 0.6-1.4 ms | ~1.6 X10$^{-2}$ A/W | 2.3X10$^8$ | 1-1.3 |
| MWCNT /Graphene[45] | BHJ | N/A | 1.5 ms | ~10$^3$ V/W | 1.5X10$^7$ | 1-1.3 |

[a] The table only lists the devices where major device parameters are available. The data that are not available are marked as N/A.
[b] BHJ: Bulk HeteroJunction.

**Table 2** summarises some major results for CNT-based photoconductors from 2003 to 2014, including both individual semiconducting SWCNTs and SWCNTs films. The progress in CNT photoconductors is due to the development of various methods of electron-hole pairs separation. The most effecient method is to form bulk heterojunctions (BHJs) by blending CNTs with certain types of polymers, such as poly [3-hexylthiophene] (P3HT).[44] The highest detectivity of $2.3 \times 10^8$ cm Hz$^{1/2}$/W in the NIR has been realized for a photoconductor based on s-SWCNT/P3HT BHJs. [44]

As early as 2003, Freitag *et al.* demonstrated the generation of photocurrent due to the electron-hole pairs separation in individual SWCNTs under an external electric field.[42] The experiment was performed on an ambipolar FET made with an individual semiconducting SWCNT (see **Fig. 3a**). (Note that even though a FET is used, the photoresponse is not due to photogating). A large photocurrent of around 30 pA was observed under a laser illumination with a wavelength range of 780-980 nm and a power of ~5.6 kW/cm$^2$, as shown in **Fig. 3c**. The nanotube FET acted as a photoconductor with an estimated EQE of ~10%. Later, numerous studies put attention on the spatially- and spectrally-resolved photocurrent of individual CNT devices to explore the electron-hole pair generation and separation in semiconducting SWCNTs [46, 47, 48, 49] and the nanophysics of individual CNT devices such as the formation of Shottky barriers at the CNT-metal electrode interfaces.[50, 51, 52] Although photoconductors based on individual CNTs are a useful platform to investigate the photophysics of nanodevices, the signal is too weak for practical photodetection due to the small absorption cross section. In order to improve responsivity, CNT photodetectors need to be extended to large area samples, such as CNT arrays or CNT films.

A photoconductor based on a single layer of horizontally-aligned SWCNTs was reported by Wang *et al*.[53] In their work, the devices consisted of arrays of aligned SWCNTs grown across a silicon trench (40 μm), with electrodes contacting the two ends of the as-grown SWCNTs. The resistance dropped by 22.86% under IR radiation with power less than 4 mW, and the measured response time was around 0.5 ms. The authors suggested that the photoresponse mechanim was the electron-hole pair separation along the CNT channels due to the applied external electric field. No responsivity or detectivity was estimated for these devices, but unfortunately, the low density of CNTs (~ 1 tube / micron) still gives a small absorption cross-section. Thus, to further improve the performance, it is necessary to develop approaches based on thicker films of CNTs where 100% of the light is absorbed, which requires a film thickness larger than 60-100 nm.

Initial reports of photodetectors based on such thin films were based on CNT networks, in which metallic and semiconducting CNTs are mixed together. The photoresponse mechanism of these detectors was predominantly thermal effects, such as the bolometric effect[54] and the photothermolelectric effect,[55] as we will discuss in later sections. However, a technique known as capacitive photocurrent spectroscopy was adopted to study the photocurrent spectrum in individual CNTs[48] and in pure CNT films,[47] which can distinguish the photoconductive effect from the thermal effect. Results showed that in CNT films electron-hole pairs can be dissociated into free carriers under large external electrical fields. Nonetheless, the portion of photoconductivity contributed by the electron-hole pairs separation is usually much smaller compared to oxygen desorption[56] and bolometric effects.[54]

One posssible way to enhance the separation of electron-hole pairs is to incorporate CNTs into certain types of polymers. Pioneering work in this field was done by Chen *et al*.[43, 57] In their work, IR photoconductors were made using composites of SWCNTs and polycarbonate (PC), in which the photoconductive effect was predominant over thermal effects. The films were put on ceramic substrates allowing good heat dissipation to inhibit the thermal effect (see **Fig. 4a**). The photoconductivity of the SWCNT-PC composite was increased by more than one order of magnitude compared to the pure SWCNT film under the same conditions. The photoconductivity enhancement was different for CNTs grown by different methods; for example, the resistance change due to IR illumination of CoMoCAT samples was larger than that of HiPco samples, and arc-discharge CNTs showed the weakest photoconductivity. This was ascribed primarily to the different aboundance of seminconducting SWCNTs, with the higher concentration of semiconducting SWCNTs leading to the larger photoconductivity.[57] The photoresponsivity of the device was around 0.4 μA/W under near-IR irradiation (1.3 μm), as shown in Fig. 4b, with a response time around 40-60 ms (see **Fig. 4c**). The experimental results were explained by the enhancement of local electric fields at the semiconducting SWCNT-polymer interface, which in turn favor the spearation of electron-hole pairs generated in the semiconducting SWCNTs provided they can diffuse to the interface.

Later studies found that other types of polymers can form heterojunctions with CNTs, such as (6,6)-phenyl-C71-butyric acid methyl ester (PC71BM),[58, 59] and poly [3-hexylthiophene] (P3HT)[44, 60]. These polymers can be blended with semiconducting nanotubes to form so-called bulk heterojunctions (BHJs),[44, 59] introducing numerous nano-heterojunctions in the whole blended film. The free electrons and holes from the separation at the junctions give the photocurrent under an applied electric field. For example, Wu *et al*. reported a semiconducting-SWCNT/P3HT BHJ photoconductor with a detectivity of $2.8 \times 10^8$ cm Hz$^{1/2}$/W, as shown in **Fig. 5**.[44] In general, the BHJ structure is convenient for application to different materials.[45] For example, the same group found that graphene could form heterojunctions on MWCNTs, with an IR detectivity ~ $1.5 \times 10^7$ cm Hz$^{1/2}$/W.

## 4. CNT Photodiodes

The development of CNT photodiodes has been approximately going through the same three phases as photoconductors: i) Individual semiconducting SWCNT photodiodes, ii) photodiodes based on semiconducting SWCNT arrays or films, and iii) photodiodes based on semiconducting-SWCNT film/polymer heterojunctions. **Table 3** lists some of the most important results in this field from the year 2003 to today. Since 2003, CNT photodiodes have evolved from individual-SWCNT photodiodes to semiconducting-SWCNT film/polymer hybrid devices recently, involving various device architectures, such as field-effect transistors (FETs),[42] p-i-n diodes,[61] barrier-free bipolar diodes (BFBD)[62] and heterojunctions.[63]

The development of CNT photodiodes tackles two key problems: i) How to construct a device to efficiently separate electron-hole pairs into free carriers and ii) how to increase the absorption cross-section of the device. As shown in **Table 3**, in recent years, significant progress has been achieved, with the responsivity increasing from ~ pA/W to ~ A/W. So far, the most successful CNT-based photodetector is the one based on SWCNT/$C_{60}$ PHJ with an EQE of 12% and a detectivity approaching ~ $10^{12}$ cm Hz$^{1/2}$/W.[63]

**Table 3** Summary of major CNT-based photoconductive and photovoltaic detectors from 2003 to 2014[a]

| Detector type | Materials | Architecture | EQE (IQE) | Response Time | Responsivity | Dectivity [cmHz$^{1/2}$/W] | Wavelength [um] |
|---|---|---|---|---|---|---|---|
| | Individual SWCNT[61] | p-i-n diode | N/A | N/A | ~0.4 pA/W | N/A | 1.5 |
| | SWCNT array[62] | BFBD[d] | ~10.4% | N/A | ~6.58 X 10$^{-2}$ A/W | 1.09 X 10$^7$ | 0.785 |
| Photo-voltaic detector | P3HT:SWCNTs /$C_{60}$[64] | PHJ | 2.3%(44%) | 7.2 ns | 0.023 A/W | 10$^{10}$-10$^{11}$ | 0.4-1.45 |
| | s-SWCNTs /$C_{60}$[63] | PHJ | 12.9%(91%) | N/A | 0.05 A/W | 0.6 X 10$^{12}$ | 1.2 |
| | s-SWCNT /GO/PC$_{71}$BM[59] | BHJ | 2.3% | 30-50 ms | 0.25 A/W | 0.9 X 10$^{12}$ | 0.86 |

[a] The table only lists the devices which have given out major device parameters.

The data that are not available are marked as N/A.
[b)] Bulk Heterojunction, [c)] Planar Heterojunction, [d)] Barrier Free Bipolar Diode

### 4.1. Photodiodes based on individual SWCNTs

Early studies showed that electron-hole pairs separation and photocurrent generation can be achieved by the built-in field in a CNT Schottky and p-n diode.[50, 61, 65] Typical values for individual CNT photodiodes are $V_{oc}$ ~ 100 mV and $I_{sc}$ ~100 pA under incident power densities of ~ 1 kW cm$^{-2}$ and FF ~ 0.4.[65] Such low sensitivity is due to extremely small absorption cross-section of individual CNT devices. In fact, the signal from individual CNTs is too weak to be useful for practical IR detection,[36] but an individual CNT diode provides an ideal platform to study the mechanisms of photocurrent generation in CNTs, such as the exciton dynamics as well as many other optoelectronic properties of CNTs.[65]

The individual semiconducting SWCNT p-i-n diode was first developed by Lee *et al.* in 2005.[61] The device used electrostatic doping by two split gates to form an n-region and p-region in adjacent parts of the same SWCNT (see sketch in **Fig. 3b**). As shown in Fig. 3d, the diode generated a photocurrent of ~ 5 pA under IR illumination with a power density ~ 20 W/cm$^2$. The FF of the CNT p-i-n diode was estimated to be in the range of 0.33-0.52, and the power conversion efficiency $\eta$ ~0.2%.

This split-gate technique has been used as an ideal tool to investigate electronic and optoelectronic properties of CNTs.[49, 66, 67] In 2013, Barkelid *et al.* systematically studied the photocurrent generation of SWCNT p-n diodes.[67] Their studies revealed that the photocurrent generation of metallic nanotubes was due to thermal effects, while the photovoltaic effect was responsible for the photocurrent generation in semiconducting nanotubes. Furthermore, recent measurements with the split gate platform indicate the coexistence of photovoltaic and photothermal effects, with their relative importance depending on the doping profile.[68]

Apart from split-gates, chemical doping and asymmetric contacts have also been used to build CNT diodes.[36, 69-71] Shim *et al.* reported a CNT p-n diode made by polymer doping, which showed a performance comparable to that obtained using the electrostatic doping approach.[69] An alternative diode structure named barrier free bipolar diode (BFBD) is based on asymmetric contacts (as illustrated in **Fig. 6a**), and was first developed by Peng *et al*.[70, 72] It consists of an intrinsic semiconducting SWCNT that is asymmetrically contacted with Pd and Sc or Y. In this diode, it is thought that Pd forms an ohmic contact with the valence band of the CNT, while Sc or Y forms an ohmic contact with the conduction band, giving rise to a band bending across the whole device channel if the channel length is less than the depletioin width at the contacts. Under illumination, free electrons and holes are formed as a result of the separation by the built-in electric field, and are then accelerated to the electrodes to yield a short-circuit photocurrent or open-circuit photovoltage (see **Fig. 6b**).[36] The device structure is relatively simple, involves only doping-free fabrication processes, and exhibits similar or better IR response compared to other types of individual CNT diodes. Peng *et al* tried to optimize photodetectors based on BEBD of individual semiconducting SWCNTs in various ways, such as studying the effects of channel length,[73] fabricating virtual contacts on the single channel[74] and coupling CNTs with plasmonic nanostructures of Au.[75]

## 4.2. Photodiodes based on SWCNT films

Similar to CNT photoconductors, the absolute absorption of an individual CNT photodiode is insufficient for most photodetector applications.[36] Such applications require the use of large-area CNT films to design more realistic photodiodes. Two problems have to be considered in order to make a SWCNT-film-based photodiode with high speed and sensitivity. First, electronic type purification of SWCNTs needs to be addressed. As will be discussed further below, the presence of metallic CNTs can render the photoresponse of CNT films thermally-dominated,[67] leading to devices with lower speed and sensitivity. So in order to inhibit thermal effects, it is critical to remove or isolate the metallic CNTs from semiconducting ones. Second, one needs to increase the efficiency of electron-hole pairs separation in bulk SWCNT films. The numerous inter-tube junctions and small-diameter CNTs in a bulk film tend to trap excitons, increase the probability of non-radiative relaxation and decrease the efficiency of free carrier collection.[30]

One strategy is to use aligned semiconducting SWCNT arrays, in which inter-tube junctions can be minimized and excitons can, in principle, easily diffusive along the SWCNT channel to electrodes. Horizontally-aligned SWCNT arrays grown on quartz wafers are a good option except for the low CNT density of ~tens of tubes per micron. With a CNT-FET structure, the metallic CNTs in the CNT array can be broken down by applying a large drain-source current along the channel, while the semiconducting SWCNTs are kept intact by electrically shutting them down using a back gate.[76, 77] (Note that while this technique works for sparse monolayer arrays, it is difficult to implement for dense or thick films due to heat dissipation to nearby CNTs that also causes them to break down.)

As mentioned in the last section, the BFBD is an ideal architecture to form individual CNT diodes. Peng *et al*. implemented the approach to SWCNT networks[76] and then to aligned SWCNT arrays (shown in **Fig. 6c**).[62] The IR photovoltaic detector based on the semiconducting SWCNT array was very successful with a much better performance than that of individual CNT devices. Under IR illumination (785 nm), the short-circuit photocurrent increased linearly with increasing incident power density, and the open-circuit photovoltage increased logarithmically with increasing light intensity, exhibiting typical photodiode behaviors, as shown in **Fig. 6d**. The optimal responsivity and the $D^*$ (based on the actual area of CNTs) were ~ $6.58 \times 10^{-2}$ A/W and ~ $1.09 \times 10^{7}$ cm Hz$^{1/2}$/W, respectively. This was the first experimental estimate of the detectivity of a CNT photonic detector, which was already larger than that of most CNT bolometers. The performance could be further improved by increasing the density of semiconducting SWCNTs in the channel: it was estimated that if the CNT density reached 60 tubes/micron, the $D^*$ of the device would be nearly one order of magnitude larger and reach ~ $10^8$ cm Hz$^{1/2}$/W.

### 4.3. Photodiodes based on planar SWCNT film-polymer heterojunctions

As discussed in the last section, BFBD detectors based on SWCNT arrays suffer from the low densities of SWCNTs in the device channel. Recently, by using appropriate polymers, a more efficient device architecture has been developed due to the great enhancement of electron-hole paris separations. This new device architecture consists of planar interfaces between semiconducting SWCNTs and a conjugated polymer[30]. As shown in the diagram of **Fig. 7a**, the principle is to construct a planar donor/acceptor type-II heterojunction between the semiconducting SWCNTs and a certain type of polymer. This heteojunction has a large band offset, which exceeds the binding energy of excitons in SWCNTs, driving spontaneous separation of electron-hole pairs at the interface between the semiconducting SWCNTs and the polymer. The electron-hole pair separation then results in charge transfer from the semiconducting SWCNTs to the polymer, creating a photocurrent.

Photodetectors based on heterojunctions between semiconducting SWCNTs and polymers have been successfully realized by several groups using a variety of charge accepting materials.[44, 58, 59, 60, 63, 64, 78] The performance of these detectors has been dramatically improved compared to the SWCNT array detector and CNT-polymer photoconductors. Among them, semiconducting SWCNT/$C_{60}$ planar heterojunction (PHJ) detectors are particularly successful and have been investigated extensively.[63, 64, 79]

Bindl *et al.* systemically investigated the separation of electron-hole pairs at the interface between semiconducting SWCNT thin films and a series of polymeric photovoltaic materials.[78] The results showed that $C_{60}$ is one of the most efficient electron accepting materials to SWCNTs. A planar type-II heterojunction was formed by vertically stacking a layer of $C_{60}$ on top of a semiconducting SWCNT thin film. Photo-generated excitons are efficiently dissociated at the interface of the SWCNT/$C_{60}$ heterojunction, and electrons are transferred from the SWCNT layer into the $C_{60}$ layer, giving an open-circuit photovoltage and short-circuit photocurrent, as shown in the **Fig. 7c, d**. The good performance of the detector relies on three important conditions: i) the SWCNTs must be isolated from each other, and in particular, the metallic CNTs need to be removed or isolated from the semiconducting CNTs;[78] ii) the thickness of the SWCNT film must be in the range of the exciton diffusion length, allowing the majority of excitons to arrive at the interface before recombining, so the optimal thickness of the SWCNT film needs to be less than 10 nm;[80] and iii) the band gap of nanotubes should be large enough, because as the band gap decreases, the driving force for electron separation at the heterointerface decreases; therefore, SWCNTs with diameter larger than 1 nm are unfavorable to form the heterojunction with $C_{60}$, implying an upper limit for the working wavelength of this type of PHJ detector.[30]

In 2009, Forrest *et al.* presented a visible to near-IR photovoltaic detector based on semiconducting-SWCNTs/$C_{60}$ PHJ.[64] The detector showed excellent performance with a $D^*$ larger than $10^{10}$ cm $Hz^{1/2}$/W in the wavelength range from 400 nm to 1400 nm, and a response time of ~ 7.2 ns. The EQE and IQE were 2.3% and 44%, respectively, at a wavelength of 1155 nm. In 2012, Bindl *et al.* developed a more successful semiconducting-SWCNT/$C_{60}$-based IR detector with a $D^*$ up to $6 \times 10^{11}$ cm $Hz^{1/2}$/W (see **Fig. 7f**).[63] The maximal EQE and IQE were increased to 12.9% and 91%, respectively. In this device, the semiconductor-enriched SWCNTs were wrapped by poly (9,9-dioctylfluorene) (PFO) as the active layer.

Later work by this group showed that the EQE of the detector could be further improved by removing the excessive PFO and by proper choice of CNT chirality. The highest EQE of ~ 34% was achieved on a photovoltaic detector based on the monochiral (7,5)-SWCNT/$C_{60}$ PHJ, in which the thickness of the SWCNT layer was around 7 nm.[81]

Further improvements in planal heterojunctions could come from engineering the CNT material to optimize photon collection and exciton separation. On the one hand, because the thickness of CNTs for complete optical absorption is on the order of several tens of nanometers, further improvement of the EQE of PHJ detector could be achieved by increasing the thickness of the SWCNT film. However, because the exciton diffusion toward the interface is limited to ~5-10 nm by poor inter-nanotube coupling, most of the excitons created in the thickner films will not diffuse to the interface. Hence, a key problem is to improve the electron-hole pair diffusion, and one possibility is to use vertically aligned SWCNTs covered by a layer of $C_{60}$, in which electron-hole pair could diffuse along the CNTs towards the heterojunction, avoiding the hopping at inter-tube junctions.[30]

## 5. CNT Bolometers

Generally, photonic detectors can offer high sensitivity and fast speed compared to thermal detectors in the visible and NIR. As the wavelength increases (and the required band gap decreases), cryogenic cooling is required to maintain the performance of photonic detectors. On the other hand, thermal detectors such as bolometers can operate in an uncooled mode within a broad wavelength range. Therefore, thermal detectors play a very important role in the MIR and longer wavelength regions (e.g., THz), where photonic detectors are less efficient.

Bolometers operate via the absorption of electromagnetic radiation and its convertion into the heat, which in turn changes the resistance of the active material according to its temperature coefficient of resistance (TCR). To assess the sensitivity of a bolometer, the responsivity is given by the following expression[82]

$$R_V = \frac{\Delta V}{\Delta P} = \frac{iR\alpha\eta}{\sqrt{G^2 + \omega^2 C^2}} = \frac{iR\alpha\eta}{G\sqrt{1+\omega^2\tau^2}}, \qquad (10)$$

where $\Delta P$ is the incident power on the active detection area, $\Delta V$ is the corresponding voltage change, $\eta$ is the absorption efficiency, $R$ is the resistance of the active material, $\alpha = (dR/dT)/R$ is the TCR, $G$ is the thermal conductance to the substrate, $C$ is the heat capacitance of the detector, and $\omega$ is the modulation frequency. The response time is given by

$$\tau = \frac{C}{G}. \qquad (11)$$

Based on Equations (10) to and (11), we can see that efficient bolometers require large optical absorption, high TCR, low heat capacity, and good thermal isolation to the environment (small $G$).

**Table 4** List of major CNT-based bolometric photodetectors with key device parameters from 2006 to 2015[a]

| Material | Measuring condition | Active area [mm$^2$] | -TCR [K$^{-1}$] | Response Time[ms] | Responsivity [V/W] | Dectivity [cm Hz$^{1/2}$/W] | Wavelength [um] |
|---|---|---|---|---|---|---|---|
| SWCNTs | Suspension (vacuum)[54] | 1.75 | ~0.1-0.7% | ~50 ms | ~10 | N/A | 0.94 |
| | Suspension (patterned Si)[82] | 0.06 | ~0.17% | 40-60 ms | ~250 | 4.5X10$^5$ | 1.3 |
| | Pixels suspension (patterned Si$_3$N$_4$)[83] | 0.02 | N/A | 10 ms | N/A | 5.5X10$^6$ | 0.8-10 |
| Aligned SWCNTs (PVP matrix) | Coating (Si)[84] | 76 | ~2.94% | ~0.94 ms | ~230 | 1.22x10$^8$ | 0.3-2 |
| Aligned SWCNTs (polystyrene matrix) | Suspension (air)[85] | 5.6-28 | ~0.82% | 150-200 ms | ~500 | N/A | 2.5-20 |
| SWCNTs (PNIPAm matrix) | Suspension (air)[86] | 0.5 | ~10% | 83 ms | 48 | N/A | 1-20 |
| MWCNTs (pristine) | Suspension (patterned Si)[87] | 0.06 | ~0.07% | ~1-2 ms | N/A | 3.3x10$^6$ | 1.3 |
| | Suspension (patterned resist)[88] | 0.055 | ~0.08% | 60 ms | ~110 | 4 X10$^6$ | 0.94 |
| | Suspension (Ag antennas)[89] | 0.04X10$^{-4}$ | 0.3% | 25 ms | ~800 | 1X10$^7$ | 10.6 |
| Aligned MWCNTs (pristine) | Suspension (air)[90] | 10 | 0.144% | 4.4ms | ~30 | N/A | 0.98 |

[a] The table only summarizes the devices which have given out major device parameters. The data that are not available are marked as N/A.

Recently, many efforts have focused on building room-temperature IR bolometers based on CNT thin films since CNTs posses excellent thermal characteristics, such as high IR absorption[54], low specific heat capacity,[91] and thermal stability.[92] However, the room temperature TCR of a pristine CNT film is normally ~0.1%/K,[54, 93] while the TCR of vanadium oxide excesses 3%/K, which is the current leading platform for uncooled bolometric detection.[34] Generally, there are two major paths to improve a CNT-bolometric detector: i) increase the TCR of CNT films and ii) reduce thermal links (decrease $G$) of the device. As far as the first aspect is concerned, many methods have been explored such as the variation of the CNT film thickness,[54] sorting the CNTs by electronic type,[33] investigating the optimal diameter distribution of CNTs,[94] and engineering the activation energy for charge transport at inter-tube junctions by embedding CNTs into a polymer matrix.[84-86, 95-97] As for the second aspect, the focal point is how to suspend the device efficiently through various techniques.[82, 88]

During the last decade, CNT-based bolometric detectors have made substantial progress and the performances of these detectors is continuously being improved. **Table 4** lists some of the most important results from 2006 to 2015, in which all device parameters were measured at room temperature. In terms of active materials, we can roughly categorize CNT-based bolometric IR detectors into three groups: i) SWCNT films,[54, 82, 83, 98] ii) SWCNT-polymer composites,[84-86, 95-97] and iii) MWCNT bundles or films.[87-90, 99, 100, 101] Except on some rare occasions, the typical response time of these devices is in the range of tens of milliseconds, and the typical dectectivity is ~$10^6$-$10^7$ cm Hz$^{1/2}$/W with working wavelengths extending from the near IR (~0.94 μm) to long-wavelength IR region (~20 μm). However, the dectectivity of current CNT-based bolometers is still 1-2 orders lower than that (~ $10^8$ cm Hz$^{1/2}$/W) of commercial bolometers.[34] In going from SWCNT films to SWCNT-polymer composites, the TCR has been improved from ~0.1%/K to ~10%/K. Unfortunately, as we shall discuss later, the increase of TCR also came at the cost of other important properties such as the response time.

### 5.1. Bolometers based on SWCNT films

Early reports regarding CNT bolometers were based mainly on SWCNTs. In 2006, the pioneering work of the Haddon group demonstrated a high IR photoresponse of suspended SWCNTs in vacuum induced by a bolometric effect,[54] as shown in **Fig. 8**. Thin films of arc-discharge CNTs with different thicknesses (40 nm - 1 μm) were held in vacuum under IR radiation with an incident power of 0.12 μW and peak wavelength of 940 nm at a temperature ~50 K. The resistance of the suspended CNT film (with 40 nm thickness) decreased by 0.7% with a response time of ~ 50 ms while the unsuspended CNT film showed no detectable resistance change. Furthermore, the values of TCR for different samples were measured, with the thinnest sample (~40 nm) showing the highest TCR change from ~ 0.7% and 2.5% in the temperature range from 330 to 100 K, but the TCR of the thicker film (~1 um) was less than 0.1% in 330 K and increased only to 0.5 % at 100 K.

According to this work, the suspension of SWCNT film is imperative to achieve good thermal insulation and therefore a sufficient IR photoresponse. Also, the thickness and operating temperature play important roles in determining the TCR. The ideal bolometer should therefore be made using perfectly suspended CNT films with thickness of tens of nanometer. Wu *et al*. reported a pristine SWCNT thin film IR bolometer with a thickness of ~ 80 nm suspended on parallel Si microchannels developed by electron-beam lithography.[82] The response time of the detector was 40-60 ms with a detectivity of ~ $4.5 \times 10^5$ cm Hz$^{1/2}$/W at room temperature. This technique makes the scalable microscopic suspension of CNT thin films possible. Recent work demonstrated a SWCNT film microbolometer array of $8 \times 8$ pixels, which were fabricated using a vertical process on a thin silicon nitride film.[83] This uncooled micorbolometer operated in the NIR and the edge of the midwave-IR band with a speed of 10 ms and detectivity of ~ $5.5 \times 10^6$ cm Hz$^{1/2}$/W.

**5.2. Bolometers based on SWCNT-polymer composites**

The TCR of CNT films is usually far less than ~1% /K at 300 K,[54] so it becomes imperative to increase the TCR in order to improve the performance of CNT-based bolometric detectors at room temperature.

For CNT films made of randomly distributed CNTs, the film resistance is dominated by transport of electrons between nearby CNTs, which is highly sensitive to the temperature.[102, 103] This type of electrical transport is commonly described in the framework of the variable range hopping (VRH) model,[86, 95, 103] according to which the resistance of a CNT film is given by

$$R = R_0 \exp(E_0/k_B T)^{\beta} \quad (12)$$

where $E_0$ is the activation energy of electron hopping between neighboring tubes, $T$ is the temperature, and $\beta$ is given by $1/(d+1)$, where $d$ is the dimensionality of the electrical transport. Accordingly, the TCR of CNT films is given by

$$TCR = \frac{dR}{dT} = -\beta \left(\frac{E_0}{k_B}\right)^{\beta} \left(\frac{1}{T}\right)^{\beta+1} \quad (13)$$

Based on Eq. (13), the TCR is proportional to the activation energy $E_0$, and is also dependent on the dimension of the system. Specifically, the lower the dimension, the higher the TCR provided that $E_0$ is larger than $k_B T$. This implies that in general, there are three possible paths to optimize the TCR of a SWCNT film:[86, 95] i) High activation energy $E_0$, which can be achieved by increasing the inter-tube distance or embedding SWCNTs into non-conductive polymer matrixes; ii) low dimensionality of electron conduction; by reducing the thickness of the SWCNT film, the electron transport behavior could become more 2D-like, and thus increase the TCR according to Eq. (13); and iii) highly semiconducting-enriched SWCNT films; because semiconducting and metallic SWCNTs have opposite signs of TCR, a film of SWCNTs containing a mixture of metallic and semiconducting CNTs will have a reduced TCR; and, the absolute value of TCR of metallic SWCNTs is usually lower than that of the semiconducting ones.[33]

Of all methods listed above, increasing the activation energy $E_0$ by using polymer matrixes has been explored most extensively. Many different non-conductive polymers have been used, including polycarbonate (PC),[57, 95] polystyrene,[85] polyaniline,[96] poly(N-isopropylacrylamide)[86, 97] and polyvinylpyrrolidone[84], and many of them indeed work very well. The TCR of SWCNT films has been increased from 0.1%/K to 10%/K using this approach.[86]

In 2008,[95] an IR bolometer based on suspended SWCNT-PC composite was reported by Aliev *et al*. The experiment showed that the resistivity of the SWCNT$_{HiPco}$ (2% wt)-PC composite increased by 4 orders of magnitude from 300 K to 10 K, while the resistivity of HiPco SWCNT bukypaper only increased by 2 orders of magnitude in the same temperature range, as shown in **Fig. 9b**. The best value of TCR was around 0.3 %/K at room temperature. The IR bolometer based on this composite operated at room temperature, with a responsivity of ~ 150V/W, but a relatively slow response time of ~1 seconds (see **Fig. 9c**). Two major reasons are responsible for the slow response time. First, the thermal conductivity between the film and the environment is low; second, a large heat capacity due to the large film thickness (tens of μm) and the mass density of the composite. Slightly increasing the SWCNT loading weight into the composite could shorten the response time but at the cost of the lower responsivity. Another solution is to decrease the film thickness, which would maintain the responsivity provided that the film is thicker than the thickness for complete optical absorption.

The highest TCR for SWCNT-polymer composites was reported by Xu *et al*.,[86] who attained 10%/K, even larger than that (3%/K) of vanadium oxide. In this work SWCNTs were incorporated into poly(N-isopropylacrylamide) (PNIPAm) to make SWCNT-PNIPAm composites. The high TCR was due to the strong enhancement of activation energy $E_0$ of CNTs due to the volume phase transition (VPT) of the polymer. The response time of the bolometer was ~80 ms, which was due to the decrease of overall heat capacity and thermal conductivity at the VPT. Unfortunately, the VPT was irreversible, depending on both temperature and humidity, and thus led to the nonlinear change of TCR and instability of the bolometer.

An alternative to CNT films made of disordered CNTs is to use films where the CNTs are macroscopically aligned. In 2012, Levitsky *et al.* reported a high midwave-IR bolometric responsivity in an aligned-SWCNT/polymer composite, which outperformed most previous reports in terms of responsivity.[85] In this work, CoMoCAT (6,5)-enriched SWCNTs were embedded into polystryrene (PS) to make free standing films with thickness in the range of 10-60 μm, and the SWCNT alignment was achieved by mechanical stretching (see **Fig. 10 a**). This bolometer showed a responsivity of ~500 V/W and a response time of ~ 200 ms. Further measurements showed that the TCR of this material was around 0.82% /K, while the randomly distributed SWCNT-PS composite and the pure SWCNT network had TCRs of ~0.4%/K and 0.17%/K, respectively.

So far, the bolometer with the highest performance was reported by Gonzalez *et al.* (see **Fig. 11**).[84] In their work, a bolometer based on a SWCNT-PVP (polyvinylpyrrolidone) composite showed a responsivity $R_V$ = 230 V/W, detectivity $D^*$ = 1.22 × 10$^8$ cm Hz$^{1/2}$/W and a response time $\tau$ = 0.42 ms. Unfortunately, the tested wavelength range (300 nm - 2 μm) was shorter than that where bolometers are typically used, so it remains to be seen if the performance can be maintained at longer wavelenghts. Most recently, the same team investigated the bolometric properties of semiconducting and metallic SWCNT composite films by using the same polymer.[33] The TCR of the semiconducting sample (95% semiconducting CNTs) has a higher value of ~6.4 %/K, while the metallic sample (95% metallic CNTs) showed a lower TCR value of ~2%/K. These results demonstrate that the type of CNTs has a direct impact on the TCR.

### 5.3. Bolometers based on MWCNT bundles or films

Because multiwall CNTs (MWCNTs) have large diameters and are composed of metallic and semiconducting SWCNT shells, they tend to show smaller values of TCR compared to SWCNTs[87], which is a disadvantage for constructing a bolometer, as shown in **Table. 4**. However, the unique structure of MWCNTs provides some advantages for IR detection. For example, it leads to an enhanced light absorption per tube, with the enhancement proportional to the number of inner shells. In addition, inner CNT shells can have a reduced thermal link to the environment and increase the sensitivity.[87] Consequently, based on existing reports, the sensitivity of MWCNT bolometers is comparable to that of SWCNTs bolometers. Furthermore, from a sample preparation point of view, it is much easier to fabricate aligned, homogeneous, large-area, and suspended MWCNT thin films with various thicknesses,[90, 104] which in turn could potentially improve the performance of a detector.

Many MWCNT bolometers have been reported, including MWCNT networks,[87, 99] vertically aligned MWCNTarrays,[101] aligned MWCNT films[90] and MWCNT bundles with antenna structures.[75] Wu and coworkers reported a room temperature bolometric response (at a wavelength of 1.3 μm) for MWCNT films suspended on patterned $SiO_2$/Si substrates.[87] The observed detectivity of ~ $3.3 \times 10^6$ cm $Hz^{1/2}$/W is seven times larger than that obtained for SWCNT films under the same conditions. The MWCNT bolometer also showed a relatively fast response time of 1-2 ms. The enhanced performance of the MWCNT bolometer was attributed to its unique microscopic configurations, where naturally suspended inner-shell CNTs provide high radiation absorbance per tube, engineered thermal link to environment, and reduced inter-tube junctions.

A polarization-sensitive IR bolometer made from highly aligned MWCNT films was reported by Jiang *et al*.[90] This aligned MWCNT film with a thickness of a few nm was made by mechanically drawing MWCNTs from a vertically-aligned MWCNT array, as shown in **Fig. 12a**. The high alignment of MWCNTs not only allowed the bolometer to detect the polarization of IR radiation, but also gave a fast response time of 1-4 ms for an active area as large as 10 mm$^2$.

Recently, Lu and coworkers developed a plasmonic-enhanced MWCNT infrared bolometer.[89] In this work, MWCNTs bundles were grown horizontally and suspended across small gaps of ~100 nm formed by silver nanoantenna arrays on a SiO$_2$/Si substrate. IR light was strongly concentrated at the nanoantenna gaps due to the plasmonic effect, and was able to efficiently heat the MWCNTs. The device operated at room temperature with a responsivity as high as 800 V/W, and a detectivity of ~1 × 10$^7$ cm Hz$^{1/2}$/W under midwave-IR radiation of 10.6 μm, demonstrating the potential of increasing the performance of MWCNT-bolometers through nanoantenna techniques.

## 6. CNT Photothermopiles

The photothermoelectric (PTE) detector (also known as photothermopile) is another type of widely-used thermal detector. Different from the bolometor, a PTE detector can, in principle, work under zero current or voltage without external power consumption. While suffering from relatively low speed and low sensitivity compared to photonic detectors, CNT-based PTE detectors show ultra-broadband response, which covers the electromagnetic spectrum from the UV, visible, infrared, and the THz region. They are advantageous in the detection of long wavelength radiation, and are particulary promising for THz detection at room temperature.

The PTE detector works based on the Seebeck effect, where the light heats the device and generates a temperature gradient along the device channel, which in turn gives rise to a photovoltage (photocurrent) accross the device. The thermovoltage is defined by

$$\Delta V = -\int S \nabla T dx \qquad (14)$$

where $S$ is the Seebeck coefficient of the active material, and $\nabla T$ is the temperature gradient in the sample channel. While $S$ is an intrinsic property of the material itself, large temperature gradients require efficient device thermal isolation from the environment, similar to a bolometric detector. Based on Eq. (14), a temperature gradient must be generated in the device, and this leads to three main device designs: in the first design, the device is symmetric, and only one of the two contacts is illuminated, leading to a temperature gradient between the two electrodes. In the second design, a p-n junction is created in the channel, and the p-n junction is illuminated, causing a temperature gradient between the p-n junction and the contacts. In the third design, the film is uniform but two different metals are used for the electrodes and the device is globally illuminated.

In the case of illumination at one contact in a symmetric device [55], the thermovoltage is

$$\Delta V = \Delta T \left( S_{CNT} - S_{metal} \right) \quad (15)$$

where $\Delta T$ is the temperature difference between the two electrodes, and $S_{metal}$ is the Seebeck coefficient of the metal electrode. This component is usually small compared to that of the active material (the CNT material in this case), and can usually be neglected for a high-performing device. In the case of the p-n junction, we have instead

$$\Delta V = \Delta T \left( S_p - S_n \right) \quad (16)$$

where $S_p$ and $S_n$ are the Seebeck coefficients for the p-type and n-type CNT material. Since the Seebeck coefficients of p-type and n-type materials have opposite signs, it is immediately clear from these two expressions that a p-n junction geometry is advantageous provided that a high Seebeck coefficient can be realized for the p-type and n-type doping, and that a similar temperare gradient can be generated. (As in the case of the bolometric detector, the device response time is given by $\tau = C/G$.)

Given the simplified situation where Johnson noise dominates the detector noise, the detectivity of a CNT PTE detector depends[105] on the thermoelectric figure of merit $ZT$ as $D^* \sim \sqrt{ZT}$ where

$$ZT = \frac{S^2 \sigma}{\kappa} \quad (17)$$

with $\kappa$ the thermal conductivity of the CNT film. Thus, improving the performance of PTE detectors requires improving the full thermoelectric properties of the active material. CNTs are suitable to make broadband PTE detectors working in the longwave-IR and the THz due to the following reasons: i) strong optical absorption due to the free carrier absorption and plasmonic resonances which can lead to large temperature increases,[27] ii) relatively large Seebeck coefficient, with values of ~ 80 µV/K to ~160 µV/K due to enhancement through the incorporation of polymers[106] or electronic type separation[107], iii) conversion from p-type to n-type by diverse doping methods,[108] allowing for the fabrication of CNT-PTE detector based on p-n junctions,[109, 110] and iv) the flexibility and strong mechanic properties,[111] allowing CNT films to be suspended or deposited on many types of substrates.[112]

**Table 5.** List of major CNT-based PTE photodetectors (also known as photothermopiles) with key devices parameters from 2006 to 2015[a]

| Materials | Architecture | Response time | Responsivity [V/W] | Detectivity [cm Hz$^{1/2}$/W] | Wavelength [µm] |
|---|---|---|---|---|---|
| SWCNT networks[106] | p-n junction | >10 s | ~1.3 V/W | N/A | 0.98 |
| SWCNT networks[110] | p-n junction | 0.034 s | ~0.9-1.8 V/W | 2X10$^6$ | 0.66-1.8 |
| Aligned SWCNT film[113] | Asymmetric electrode | ~32 us | ~0.028V/W | 1.1X10$^5$ | 0.66-3.3 |
| Aligned SWCNT film[109] | p-n junction | 0.1-0.6 s | ~1 V/W | N/A | 0.8-3.2 |
| SWCNT networks[114] | Asymmetric thermal contact | N/A | ~2.5V/W | N/A | 140 GHZ |
| DWCNT network[115] | Asymmetric thermal contact | ~0.2-8 s | ~0.022 V/W | N/A | 2.52 THZ |
| Aligned SWCNT film[112] | p-n junction | ~ 0.1s | ~ 2.5 V/W | 5×10$^6$ | 1.39-3.11 Hz |

[a] The table only summarizes the devices which have given out major device parameters. The data that are not available are marked as N/A.

CNT-PTE detectors have been made using large-size CNT films, including CNT networks,[106, 110, 116] aligned CNT films[109, 112, 113] and CNT-polymer composites.[117] The device structures are diverse, including p-n junctions,[106, 109, 110, 112] asymmetric metal electrodes [113]and asymmetric thermal contacts.[114, 115] **Table 5** lists the characteristics of some recent reports of CNT-PTE detectors. As can be seen, the response time of this type of detector is normally slow, in the range from milliseconds to seconds, due to the reliance on thermal dissipation. Currently, the detectivity of CNT-PTE detectors is also low, less than 10$^7$ cm Hz$^{1/2}$/W, which is comparable to most CNT bolometers but much below the best photonic detectors in the NIR. However, as mentioned before, the advantage of these detectors is their broadband response which allows their application in regimes where photonic detectors have generally poor responsivity. A particular example is the THz regime, which is notoriously difficult for photonic devices. There are already two prototypes of THz PTE detectors based on CNT films,[112, 115] as shown in **Table 5**. Possible paths to further improve the speed and the sensitivity of CNT PTE detectors include, i) Improving the thermoelectric properties by exploring the CNT alignment, density, diameter distribution, quality of junctions, etc, ii) Developing approaches to make suspended ultrathin films in order to reduce the heat capacity and improve the response time, and iii) Developing approaches for voltage multiplication based on multiple junctions.

**6.1. PTE effect in carbon nanotubes**

A detailed understanding of the photocurrent generation mechanism in carbon nanotubes is the first step towards optimizing the design of CNT-based photodetectors. In principle, the photovoltaic and PTE effect coexist in CNTs based photodetectors, and the relative importance of one or the other mechanism is strongly dependent on the electronic types and assembly of CNTs, as well as the device configuration.[68]

Individual CNT diodes have been investigated intensely using spectrally and spatially resolved photoconductivity. As discussed in previous section, studies suggested the photovoltaic effect was responsible for the photocurrent generation for substrate-bound individual s-SWCNT device, where the sample heating due to illumination was strongly inhibited. The importance of PTE effect was revealed for suspended SWCNTs with no and small band gaps.[52, 67, 118] Very recently, the question of the role of PTE effect in large bandgap semiconducting nanotubes has been studied in suspended CNT devices.[68] In this work, Scanning Photocurrent Microscopy (SPCM) was used to investigate the local photocurrent generation of double-gated suspended s-SWCNT photodiodes (as illustrated in **Fig. 13a**). The 2D map of the photocurrent as a function of electrostatic doping showed a fourfold-like pattern (see **Fig. 13b**), indicating the coexistence of both photovoltaic and photothermal mechanisms in the device. Detailed analysis revealed that photocurrents generated by the PTE effect and the photovoltaic effect are in the same direction under the p-n doping profile, but in opposite directions under the p-p⁻ and p-p+ doping profiles. In this case, the PTE effect could be dominant and reverse the sign of the photocurrent, if the contact resistance at the CNT-electrode interface was not very high. These results show that the PTE effect is generally present in CNTs regardless of the electronic type.

The first example of a macroscopic PTE effect of bulk SWCNT film was presented by the Martel group for symmetric devices with contact illumination.[55, 119] Under local illumination of macroscopic films, previous studies had observed a sign change of the unbiased photovoltage (photocurrent) from one side of the device channel to the other, similar to the photovoltaic effect of individual semiconducting SWCNT except for the much slower response. The position effect initially was explained as originating from the diffusion of photoexcited carriers in the band-bending potential due to Schottky barriers at the nanotube-electrode junctions.[120] In 2009, Martel *et al.* investigated the unbiased photovoltage on suspended CNT films,[119] demonstrating that the photovoltage originated from light-induced heating and the spatial variation of the Seebeck coefficient along the CNT film. Later, the same team demonstrated that the PTE effect was even dominant on macroscopic CNT films deposited on substrates.[55] Two important features were: i) The photovoltage generated at the electrode/ CNT interface was very sensitive to the substrate. With a substrate of high thermal conductivity (sapphire), the photovoltage magnitude was strongly suppressed, but it was enhanced on the substrate with low thermal conductivity (glass), as shown in **Fig. 14a**. On the contrary, the response time became much slower on the substrate with low the thermal conductivity (glass) (see **Fig. 14b**). The inherent link between speed and sensitivity with the substrate thermal conductivity is a typical behavior of PTE detectors. ii) The photovoltage could be modified by using different metals for the electrodes. The difference of the Seebeck coefficients between the electrode material and CNTs rather than the work function difference between them determined the photoresponse at the interface, indicating that the device behaved like a thermocouple in this situation. The results are shown in **Fig. 14c**, where the largest photoresponse came from the interface of SWCNT-bismuth electordes as it gave the largest difference of Seebeck cofficients (the Seebeck cofficient of the p-type SWCNT film was ~ 30 µV/K).

**6.2. PTE detectors based on carbon nanotubes**

Most CNT-PTE detectors are based on large size CNT films, including CNT networks,[106, 110, 116] aligned CNT films [109, 112, 113] and CNT-polymer composites.[117] These devices show braodband photoresponse spanning from the IR to the THz with different device structures such as p-n thermal junctions, metal-CNT junctions, and asymmetric thermal contacts.

In 2010,[110] the Martel group designed a thermopile based on a p-n doping profile in a suspended SWCNT film, as shown in **Fig. 15a**, **b**. The p-n doping profile was realized by depositing potassium on part of the SWCNT film. While the original p-type CNT film had a Seebeck coefficient of ~30 µV/K, for the potassium-exposed region the Seebeck coefficient was ~ −10 µV/K (see the **Fig. 15c**). The CNT thermopile showed a time response of 36 ms, maximal responsivity of 1.6 V/W( see **Fig. 15d**),and a detectivity of $2 \times 10^6$ cm Hz$^{1/2}$/W in the visible and near-infrared. This work was important in demonstrating that the SWCNT-based thermopile could be used as a broadband light detector with the overall performance comparable to that of SWCNT bolometers.

Fan *et al*. successfully turned a p-type CNT film into a n-type one by using a more efficient doping method based on polyethyleneimine (PEI).[106] SWCNT sheets displayed a large negative coefficient (~-87 uV/K) after PEI doping, while the original p-type sheet showed a positive coefficient (~ 70 µV/K). By integrating p-type and n-type CNT sheets in series, a NIR opto-electronic power source was constructed, in which the connection of each p-type and n-type CNTs forms an individual thermocouple (as illustrated in **Fig. 16 a**). The output photovoltage of the device showed a good linear relationship with NIR light power (985 nm), and under global illumination, the device (consisting of 50 p-type elements and 50 n-type elements) had a responsivity of ~ 1.3 V/W (see Fig. 16 b), and a response time longer than 10 s. When compared to previous work with a single thermopile element based on CNT thin films, the response time of this device is much slower, and the responsivity is of the same order. The reason is possibly due to the much larger thickness of CNT sheets (~20 µm, and therefore a large heat capacity), and the global illumination which could have led to small temperature gradients. Nevertheless, the responsivity of the device with p-n junctions in series was around two orders larger than a single p-type (or n-type) element made with CNT sheets of the same thickness, indicating that multiple p-n junctions indeed could boost the sensitivity of the device.

Because the above devices were based on disordered networks of CNTs, one question is whether the degree of order in the film plays a role in the performance, and if it can lead to new functionality (such as polarization sensitivity). The first CNT-PTE detector based on aligned CNT film was realized by our group.[113] In this work, horizontal SWCNT thin films were made by dry transferring fins of vertically aligned, ultralong SWCNTs onto SiO$_2$/Si substrates, as shown in **Fig. 17**. The length and the thickness of the film were 300 µm and 0.6 µm respectively. The IR response (0.66 to 3.15 µm) of the aligned SWCNT film was examined using SPCM with electrodes made by different metal materials (Au, Pd and Ti).

Unbiased photovoltages and photocurrent were observed at the interface of the electrodes and SWCNT channel with sign change from one side to the other, similar to previous results from CNT networks. The mechanism of the photocurrent was identified as the PTE effect by detailed experimental and theoretical analysis. The largest photoresponse was obtained at the interface of Ti electrodes and the SWCNT film, which was due to the highest increase in temperature compared to devices with other types of electrodes. A PTE detector was made by forming asymmetric electrodes at the two ends of the aligned SWCNT film (Pd and Ti). The device generated net photocurrent under global illumination with a responsivity of ~ 0.028 V/W, which was lower than the PTE detector based on CNT networks. There were a few reasons for the low responsivity: i) Global illumination led to weak temperature increase; ii) The sample was on a substrate with good thermal conductivity ($SiO_2$) instead of being suspended. The low responsitvity is balanced by the exceptionally high speed of the detector (~32 μs) compared to other thermal detectors, as shown in **Fig. 17d**. Interestingly, the aligned CNT-PTE detector showed a unique and useful feature— polarization sensitivity. A photorespone ratio of ~0.5 was observed as the light polarization was rotated between the parallel and perpendicular direction of the CNT alignment, as shown in **Fig. 17c**.

To improve the sensitivity of the aligned CNT-PTE detector, we later designed a new device based on a p-n junction. The CNT thin film was formed by rolling down SWCNT vertical arrays.[109] Because of the p-n junction, the active area of the device was moved to the center of the channel, where the p-n thermal junction was made by overlapping one p-type SWCNT film with another n-type film (see **Fig. 18a**). The n-type doping was achieved using benzyl viologen (BV), which is an effective and air-stable n-doping method. Based on **Fig. 18b**, the responsivity of the device supported by Teflon tape was around ~ 1V/W, comparable to other CNT-PTE detector based on p-n thermal junction. A systematic analysis of the role of the substrate revealed the trade-off between the responsivity and response time of the detector.

More importantly, this device was the first p-n junction CNT thermopile that showed polarization sensivity, and also extended the demonstration of photoresponse to the MWIR with sustained efficiency. This demonstration paved the way to further exciting studies in the THz range.

A CNT-PTE detector can work in the longwave-IR due to free carrier absorption, which in principle extends to the THz region. CNT-PTE detectors showing response under sub-THz radiation were made by Goltsman *et al*.[114] In these devices, CVD-grown CNT films were partially supported by catalyst islands and partially contacted to bottom $SiO_2$ substrate, leading to an asymmetric thermal contact to the substrate (see **Fig. 19a** and **b**). When exposed to radiation, a temperature gradient was generated along the CNT film due to the difference of thermal coupling to the substrate. A spiral antenna was attached to the CNT film to collect the radiation efficiently. These devices showed a DC voltage signal in response to radiation of 140 GHz in the temperature range from 4.2 to 300 K, as shown **Fig. 19c**. The best room temperature responsivity of the device was around ~2.5V/W, and it reached ~500 V/W at low temperature (4.2 K).

Recently, we developed an antenna-free CNT-PTE detector working in the THz region, as shown in **Fig. 20**.[112] This detector was fabricated on a single piece of aligned SWCNT film grown by a CVD method, with a film thickness around 1-2 μm. The width of the SWCNT film and the device channel length were 150 um and 1 mm respectively. A p-n thermal junction was made by partially BV n-doping the as-grown p-type film. The whole device was supported on suspended Teflon tape for good thermal isolation. A photovoltage was observed under the THz radiation of 1 to 3 THz, with the average response of ~2.5 V/W, as shown in **Fig. 20d**. Thermoelectric measurements showed a Seebeck coefficient for the original p-type film of +70 μV/K, while that of the n-type film was -70 μV/K. In addition, the detector was polarization sensitive due to the well-aligned structure of the SWCNT film, with the ratio of photovoltage between parallel and perpendicular polarization between 0.1 to 0.3 (see **Fig. 20c**), much larger than that in IR region. The results open the possibility of making uncooled THz detectors based on macroscopic CNT films. However, these devices still suffer from limited active area under THz radiation, low speed on the order of seconds, and low sensitivity. Further improvement is expected by: i) increasing the active area by the fabrication of multiple junctions on the same devices, ii) using SWCNT films with high semiconducting enrichment to boost the Seebeck coefficients and therefore the responsivity.

## 7. Conclusions

In conclusion, the field of CNT photodetectors has evolved tremendously over the past few years. From the initial scientific demonstrations of photocurrent in individual CNTs, a number of macroscopic devices relevant to applications from the visible to the THz have since emerged. Fundamental studies have now clearly established the different mechanisms that can govern these detectors, and the paths towards improving their performance. We summarize the state of the field in **Fig. 21**, where we show the detectivity of various types of devices in the visible, NIR, midwave-IR, and THz regimes. It is clear from this plot that most of the work has focused on visible or NIR detectors, with the photonic detectors showing the best performance. Much less work has been presented in the midwave-IR and THz range, where currently thermal CNT detectors are the only demonstrated devices.

In the NIR range, the best existing uncooled detectors (e.g., InGaAs) attain detectivities of $10^{13}$ cm $Hz^{1/2}$/W, so the demonstration of a CNT photovoltaic device with detectivity of $10^{12}$ cm $Hz^{1/2}$/W is quite remarkable. In the MWIR range, the performance of existing detectors (e.g., PbSe) drops to $10^9$-$10^{10}$ cm $Hz^{1/2}$/W; despite this reduction in performance, the existing CNT detectors are further from existing technology than in the NIR case, in part because much less work has been done in this regime. The situation is quite encouraging in the THz because of the relative lack of existing technology: typical[121] uncooled THz detectors (e.g., microbolometers) attain detectivities of $10^9$ cm $Hz^{1/2}$/W, so it is quite promising to see that the few CNT photodetectors demonstrated in this regime are already in reach of this value, particularly since the potential and path towards improvement have been clearly established.

**Acknowledgements**

This work was supported by the US Department of Energy, Office of Science under the National Institute for Nano Engineering (NINE) at Sandia National Laboratories, and by the Lockheed-Martin Rice University LANCER Program. X.H. and J.K. were supported by DOE BES DE-FG02-06ER46308 (preparation and teraherz/infrared characterization of aligned carbon nanotubes) and the Robert A. Welch Foundation Grant No. C-1509 (detector fabrication). Sandia National Laboratories is a multi-program laboratory managed and operated by Sandia Corporation, a wholly owned subsidiary of Lockheed Martin Corporation, for the U.S. Department of Energy's National Nuclear Security Administration under contract DE-AC04-94AL85000.

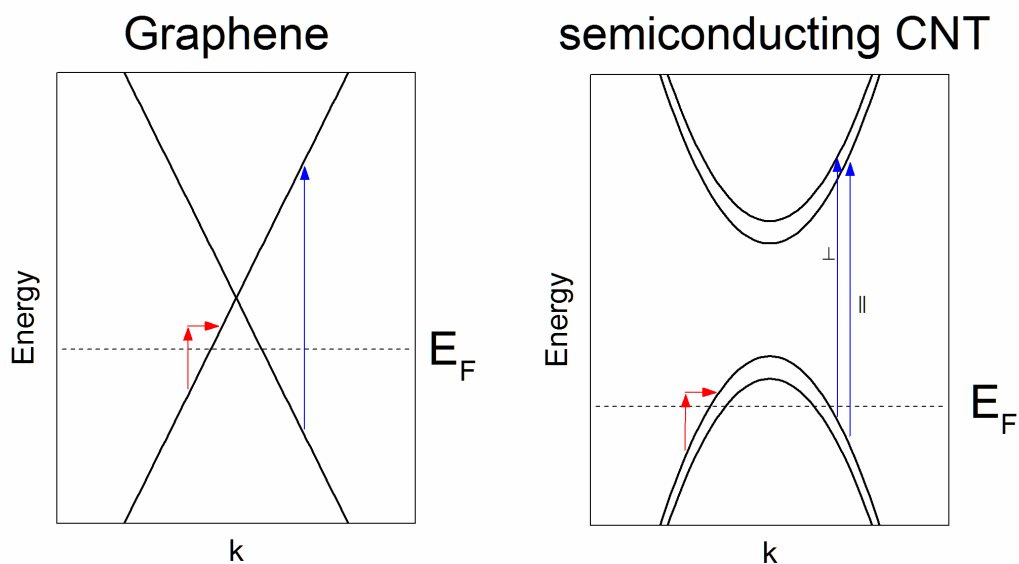

**Figure 1.** Illustration of the optical absorption processes in graphene (a) and carbon nanotubes (b) in terms of the electronic bandstructure. In both cases intraband transitions are illustrated with red arrows while interband transitions are shown with blue arrows.

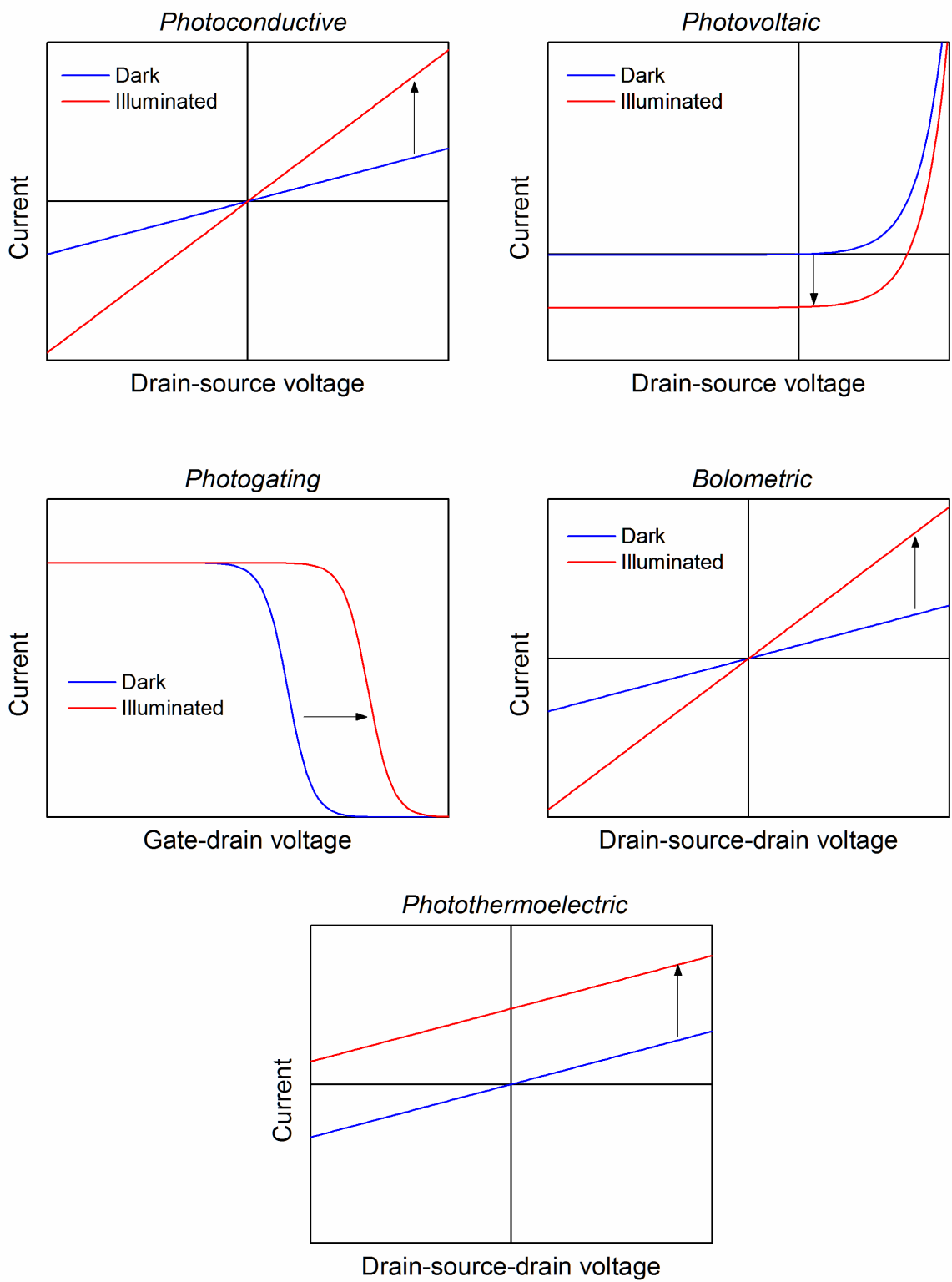

**Figure 2.** Current-voltage characteristics of the five types of photodetectors in the dark and under illumination.

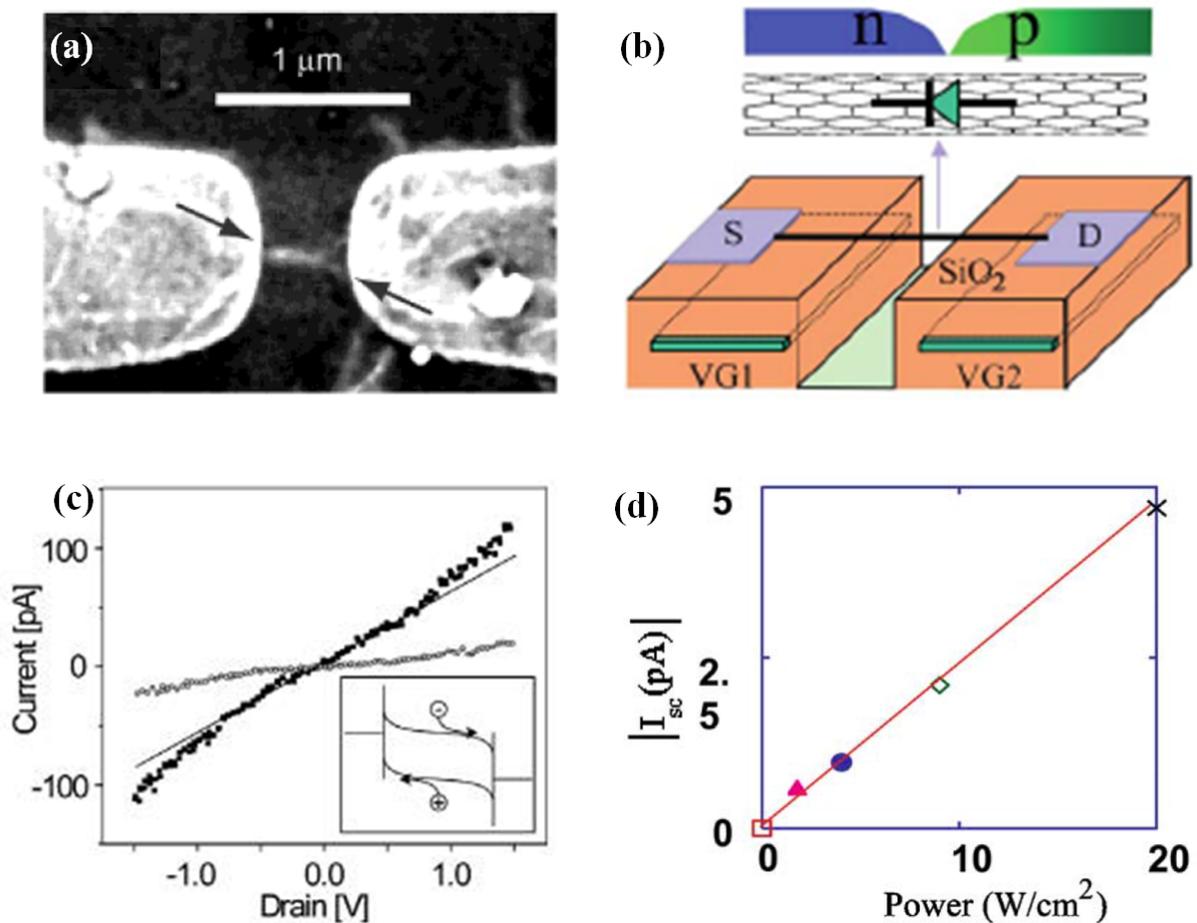

**Figure 3.** a) SEM image of a photoconductor based on the individual SWCNT FET. b) Schematic diagram of individual SWCNT photodiode with split gates. c) Drain-Voltage dependence of photocurrent on the s-SWCNT FET based photoconductor without light and (o) with infrared light (•). d)The open circuit photocurrent of individual SWCNT photodiode as a function of illuminated power. Figure 3 a) and c) are adapted with permission.[42] Copyright 2003, American Chemical Society. Figure 3 b) and d) are adapted with permission.[61] Copyright 2005, AIP Publishing LLC.

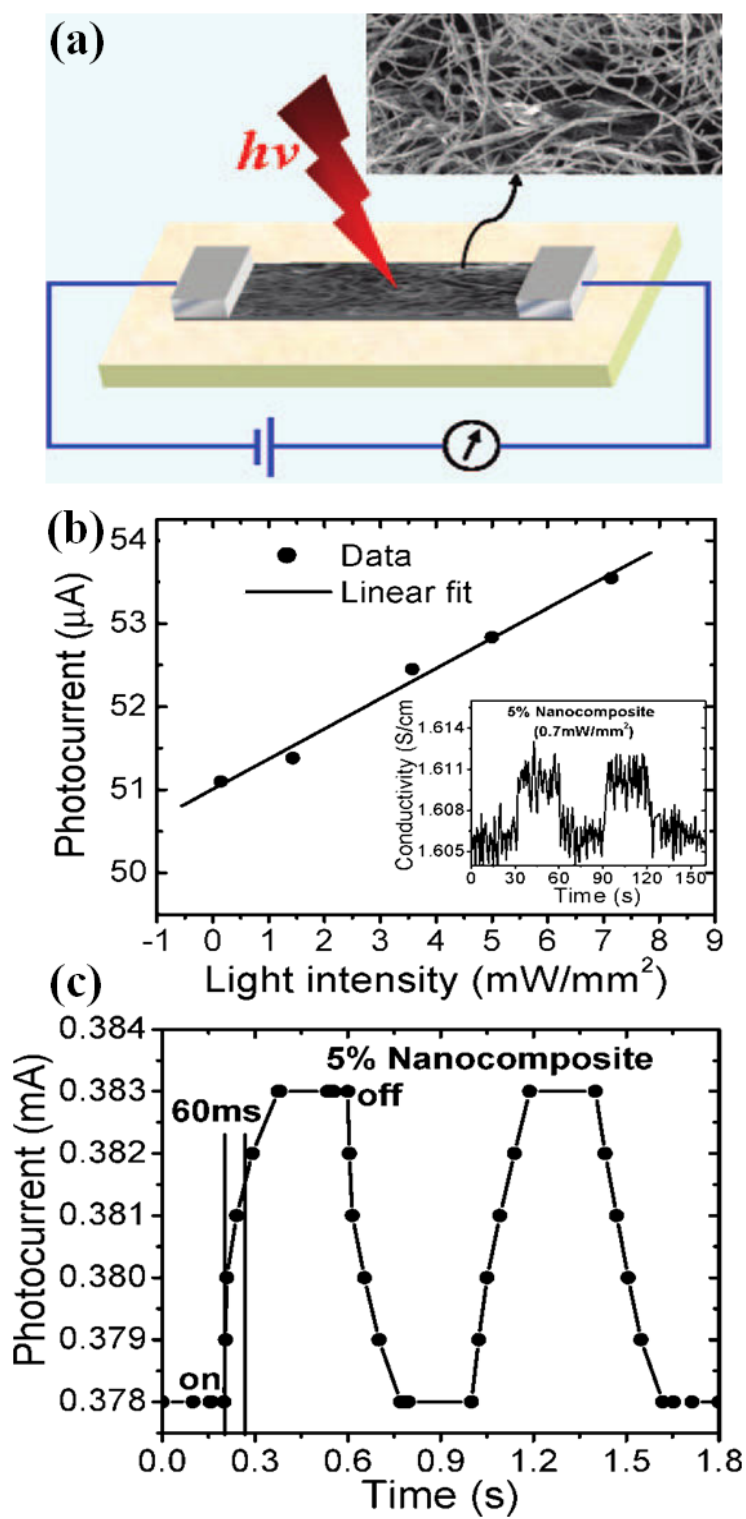

**Figure 4.** a) Schematic map of the photoconductor based on SWCNT$_{Hipco}$-PC composite film, (inset is the SEM image of the film). b) Dependence of photocurrent on the IR light intensity. Inset: photoconductivity response to the on/off illumination at 0.7 mW/mm$^2$ light intensity. c) Response time of the photocurrent to IR illumination (power intensity: 7mW/mm$^2$). Reproduced with permission.[43] Copyright 2008, American Chemical Society.

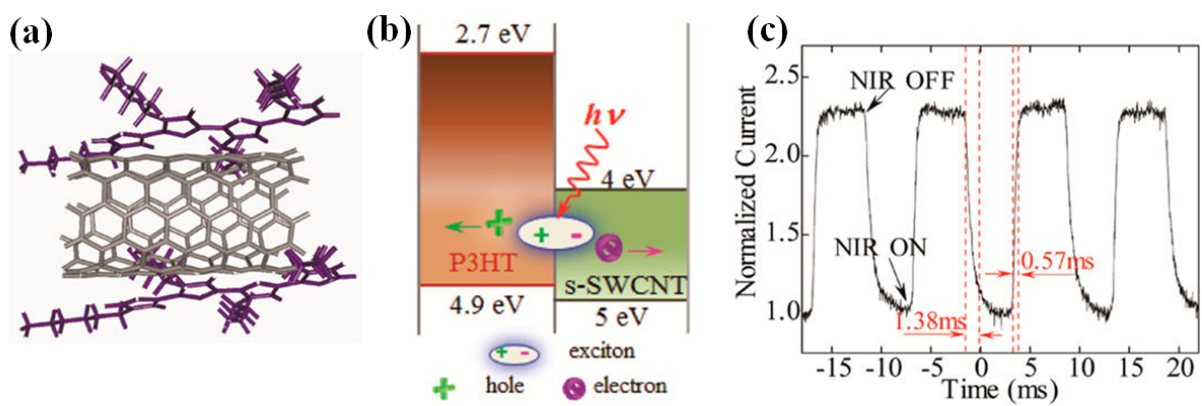

**Figure 5.** a) Diagram of s-SWCNT/P3HT nanohybrid. b) Band structure of the s-SWCNT/P3HT type-II heteojunction. c) Photocurrent response of the SWCNT/P3HT photoconductor under 99 Hz NIR modulation with the intensity of 0.35 mW/mm$^2$ and bias voltage was 10 V. Reproduced with permission.[44] Copyright 2012, American Chemical Society.

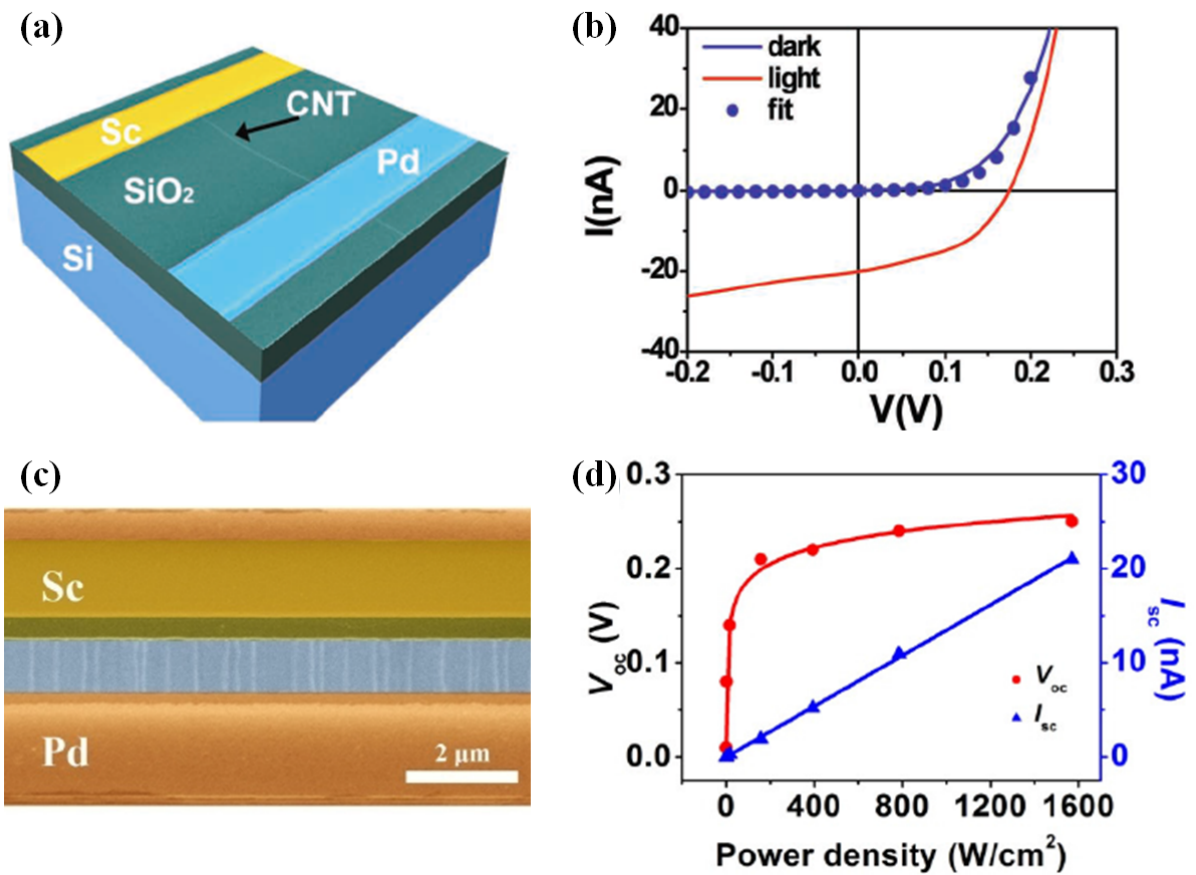

**Figure 6.** a) Schematic image of a barrier free bipolar diode (BFBD) based on a single SWCNT. b) Experimental (solid blue line) and fitted (blue dots) the CNT diode characteristic in dark, and experimentally measured (solid red line) I-V characteristic under illumination. c) SEM image of photovoltaic detector based on SWCNT array BFBD, with channel width W= 20 μm and channel length L =1 μm. d) Experimental data and fitted results for open circuit photovoltage and short circuit photocurrent as a function of illumination power density on the SWCNTs photovoltaic detector . Figure 6 a) and b) are reperoduced with permission.[36] Copyright 2013, John Wiley and Sons. Figure 6 c) and d) are reproduced with permission.[62] Copyright 2012, Optical Society of America.

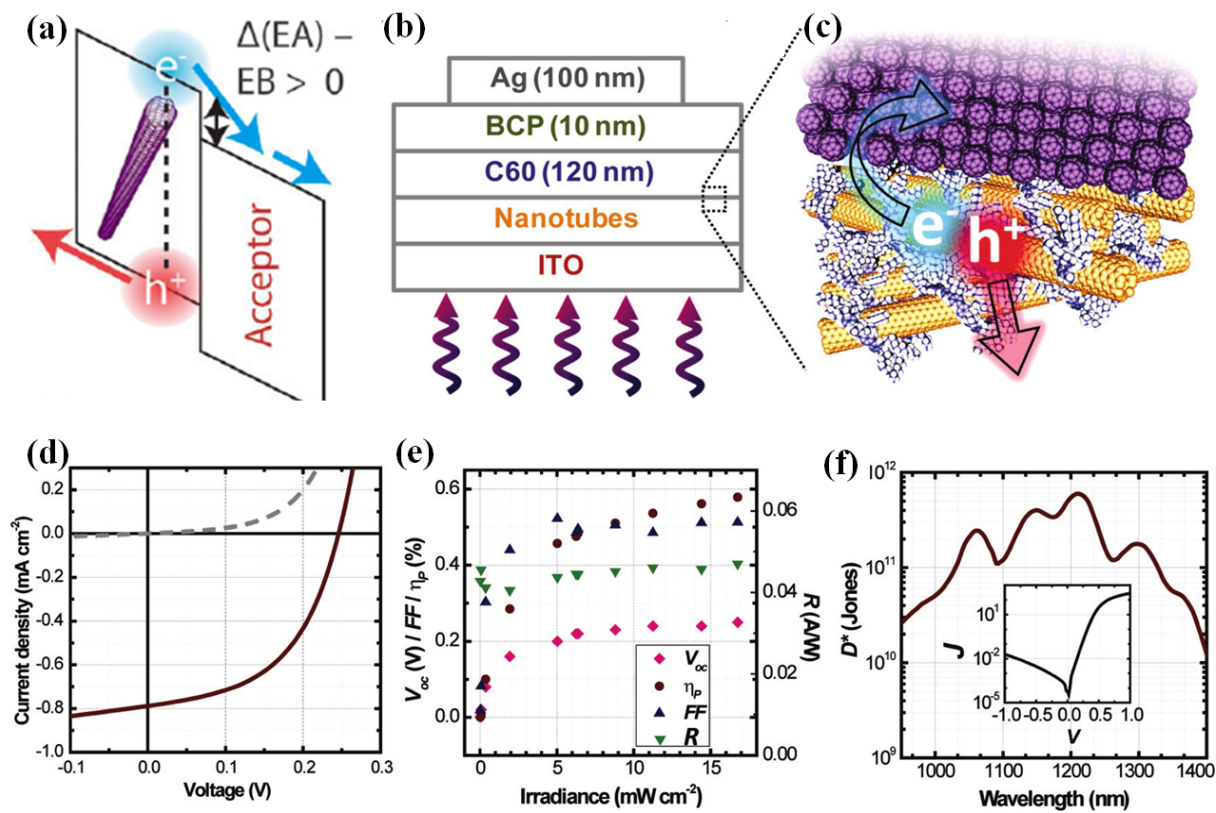

**Figure 7.** a) Energy level of a type-II donor-acceptor heteojunction for the CNT/polymer based photovoltaic detector. b) and c) Architecture of a SWCNTs-$C_{60}$ planar heterojunction device. d) I-V curves in the dark (dashed, gray) and in the illumination of NIR (1-1.365 μm) with power density of ~17 mW cm$^{-2}$. e) Intesnity dependent photovoltaic parameters. f) The specific detectivity (D*) of the SWCNTs-$C_{60}$ photovoltaic detector. Inset shows the dark I-V curve of the detector (J is of units mA and V is of units Volts). Reproduced with permission.[63] Copyright 2010, American Chemical Society.

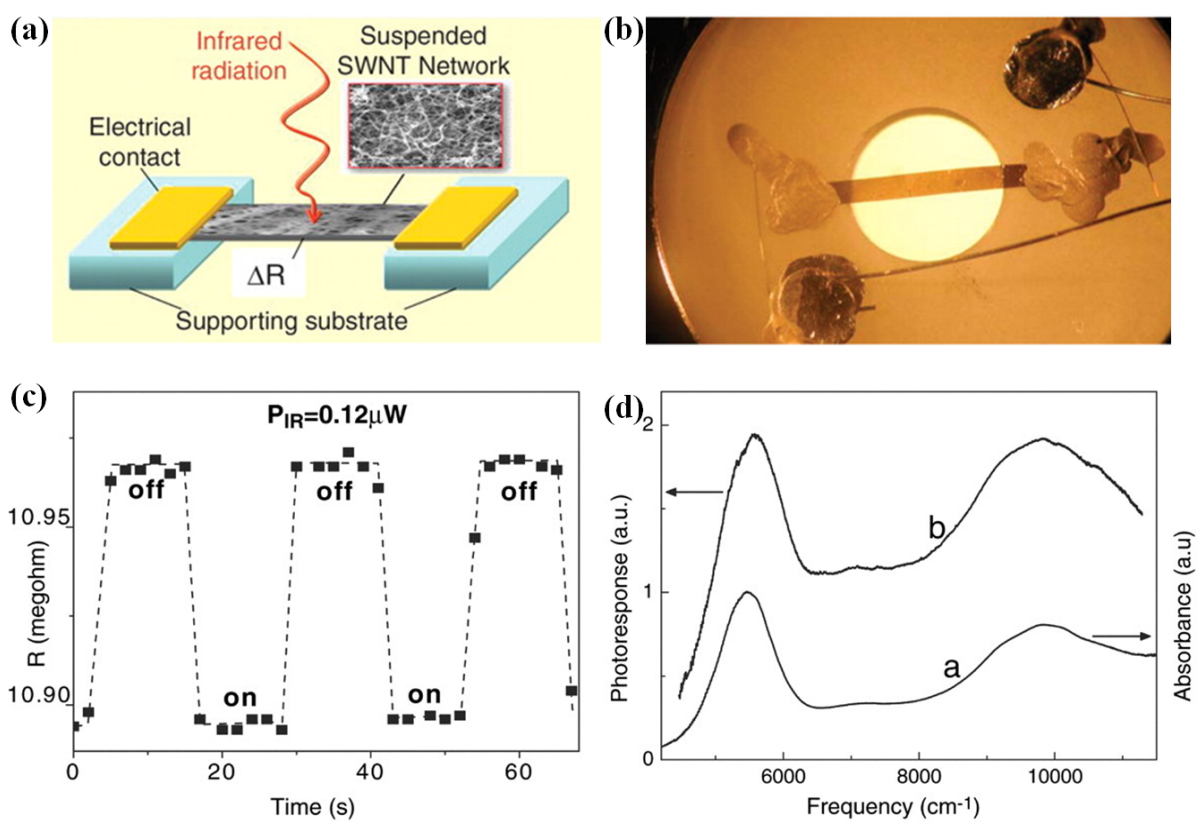

**Figure 8.** a) Diagram of a bolometer of suspended SWCNT network. b) 100-nm-thick SWCNT film suspended across 3.5 mm opening a sapphire ring. c) Modulation of resistance of SWCNT film at 50 K under square-wave pulse of IR radiation with the power of 0.12 µW. d) Spectra of near-IR absorption (curve a) and electrical photoresponse of SWCNT film (curve b). Reproduced with permission.[54] Copyright 2006, The American Association for the Advancement of Science.

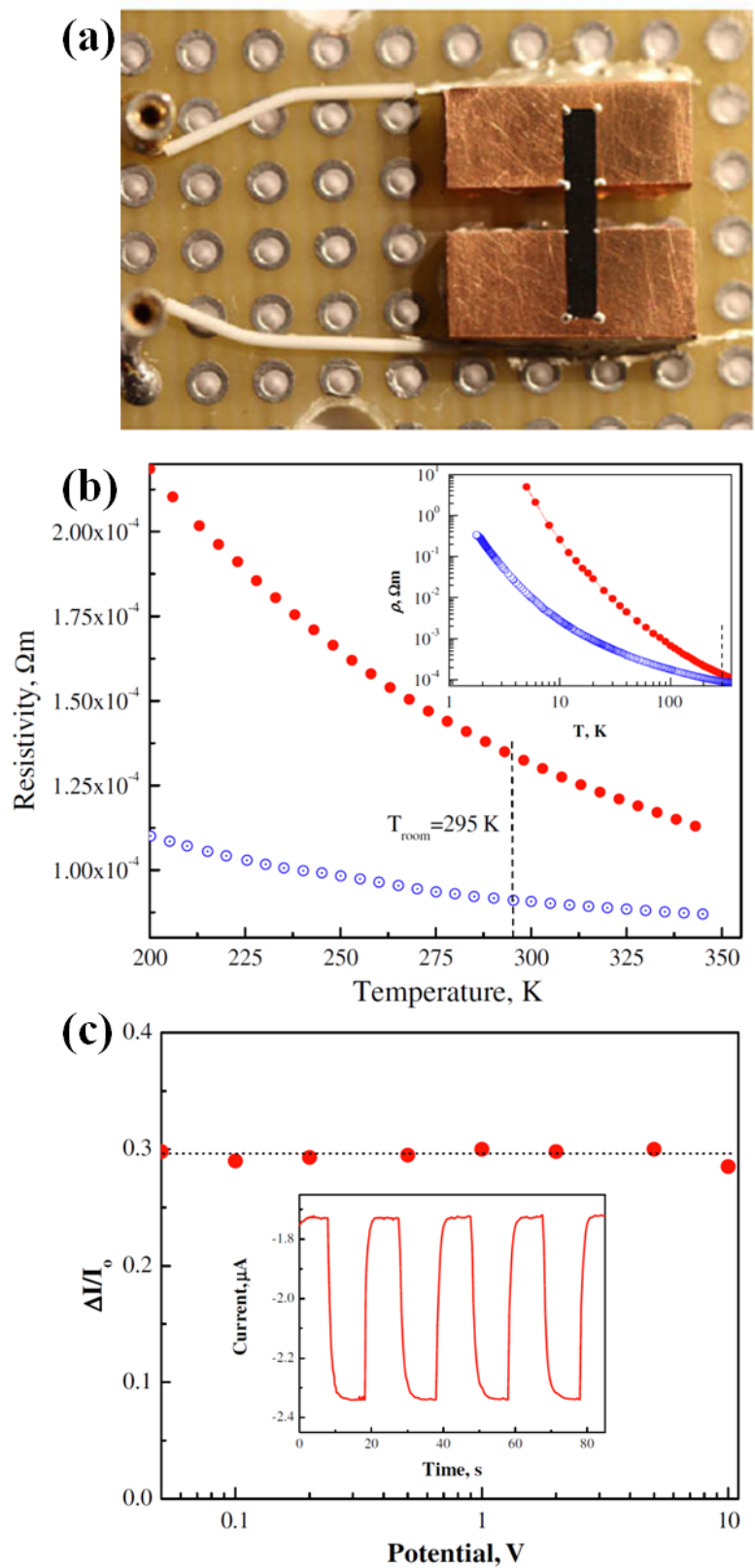

**Figure 9.** a) Bolometric detector based SWCNT/Polycarbonate (PC)composite film (9×1 mm$^2$ ) attached to two copper bricks. b) 100-nm-thick SWCNT film suspended across 3.5 mm opening a sapphire ring. c) Photocurrent modulation versus applied voltage in the SWNT/PC

bolometer, Inset shows the response time of the bolometer to 10 mW exposure by white light. Reproduced with permission.[95] Copyright 2008, Elsevier.

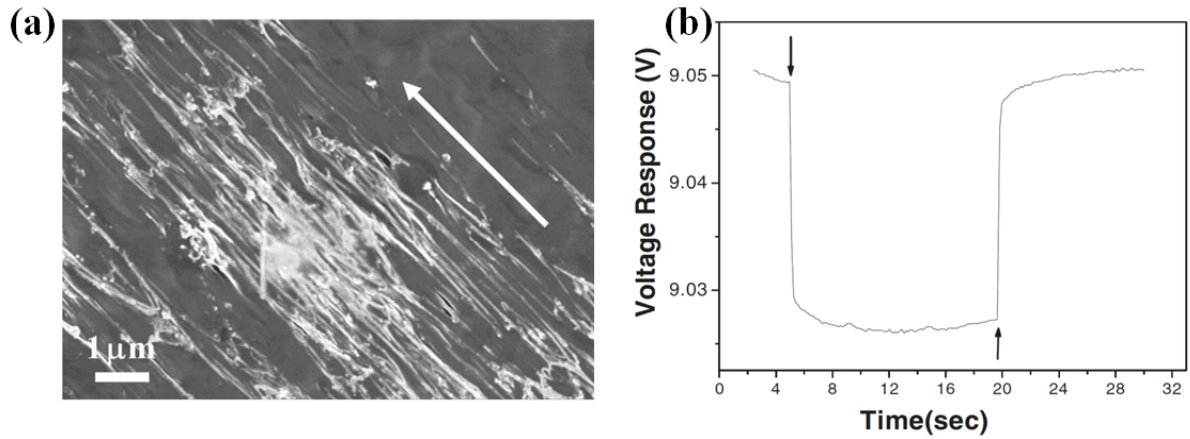

**Figure 10.** a) SEM image of SWNT-polystyrene composite stretched by thermal pulling at moderate magnification. b) Photovoltage response of the bolometer based on SWCNT-polystyrene film which has a responsivity of 21.5 V/W and a response time of 180 ms. Reproduced with permission.[85] Copyright 2012, Wiley.

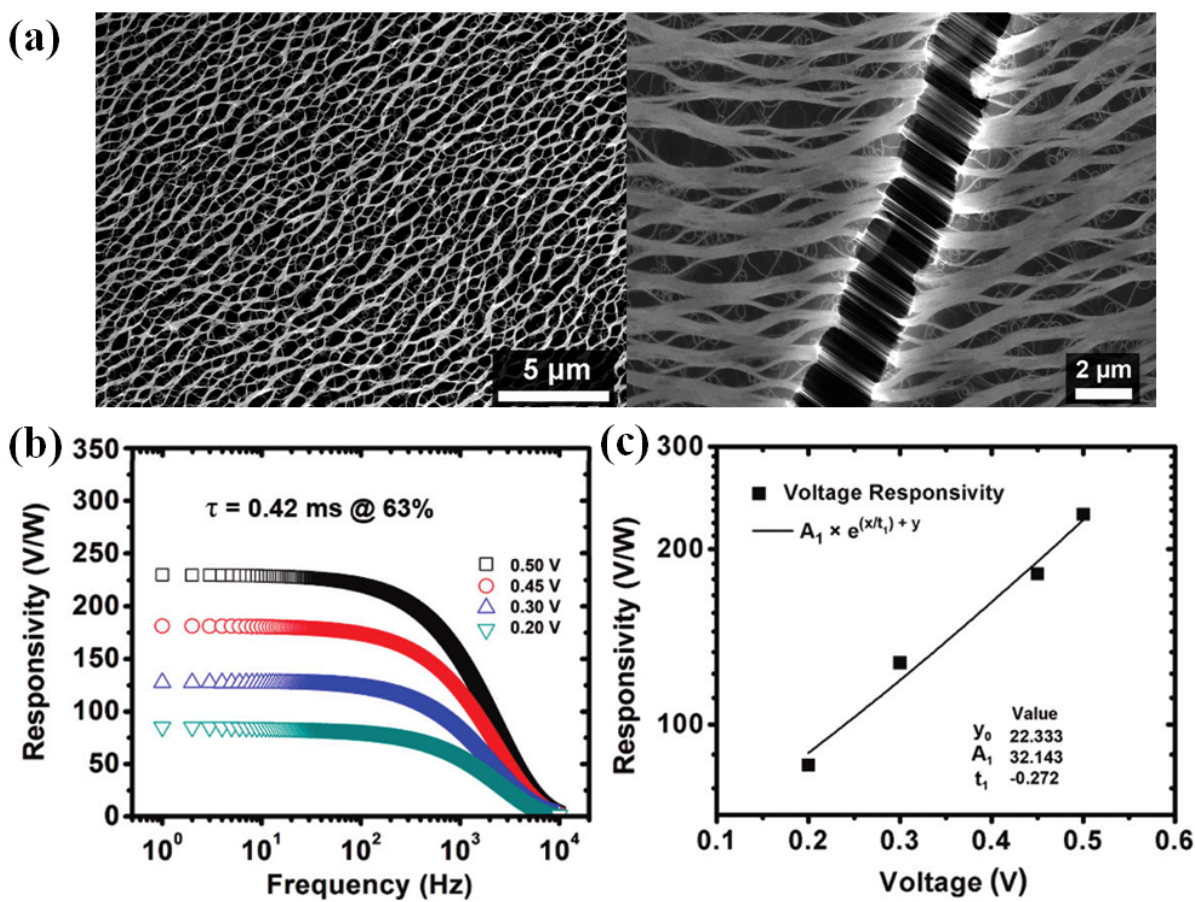

**Figure 11.** a) SEM image of horizontally aligned SWCNTs forming dense networks in the polyvinylpyrrolidone (PVP-10) matrixes. b) Frequency dependence of responsivity of the bolometer based on SWCNT/PVP-10 composites , c) Maximum responsivity as a function of bias voltage. Reproduced with permission.[84] Copyright 2011, American Cheimcal Society.

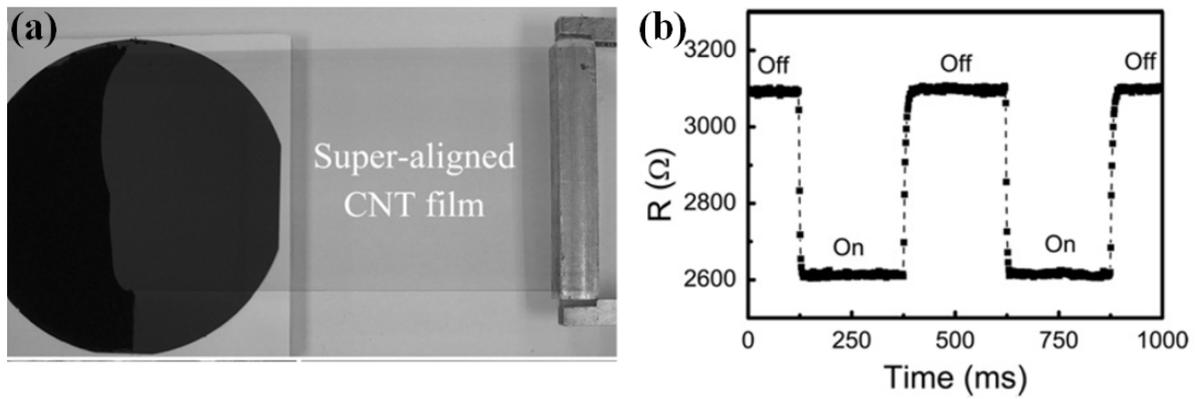

**Figure 12.** a) Super-aligned CNT film drawn from MWCNT arrays on a 4 inch silicon wafer. b)The on-off photoresponse of a bolometer based on one layer of super aligned MWCNT film under an IR radiation of 10 mW/mm$^2$ with the polarization direction of incident beam parallel to the alignment direction of CNT film in vacuum at room temperature. Reproduced with permission.[90] Copyright 2010, IOP publishing.

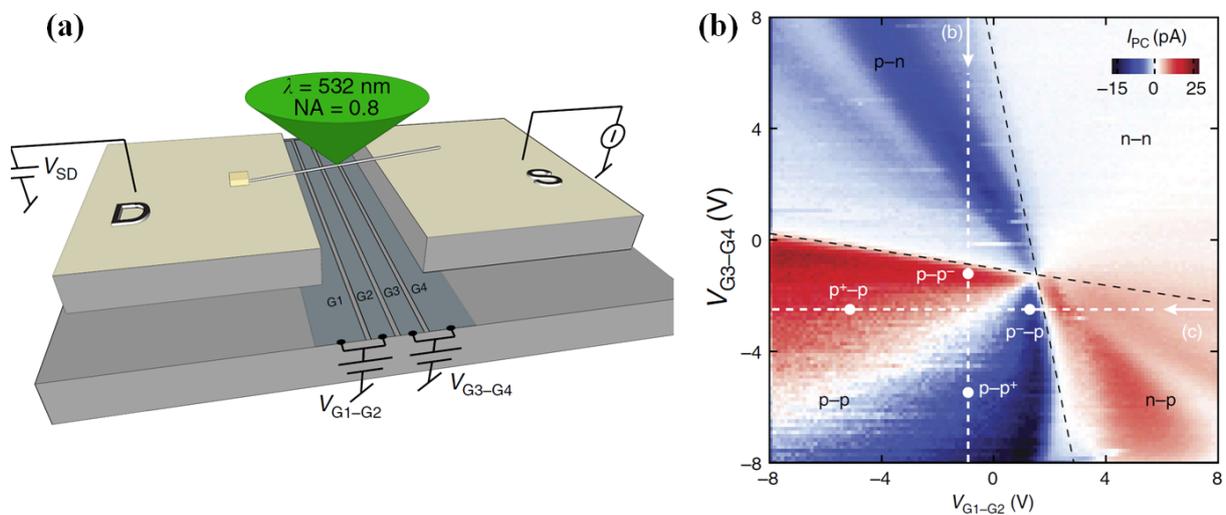

**Figure 13.** a) Schematic of single s-SWCNT FET suspended on a trench with split back gates. b) A 2D map of the photocurrent as the function of doping profile generated by back gates with the laser focused near the middle section of the suspended nanotube. Reproduced with permission.[68] Copyright 2014, Nature Publishing Group.

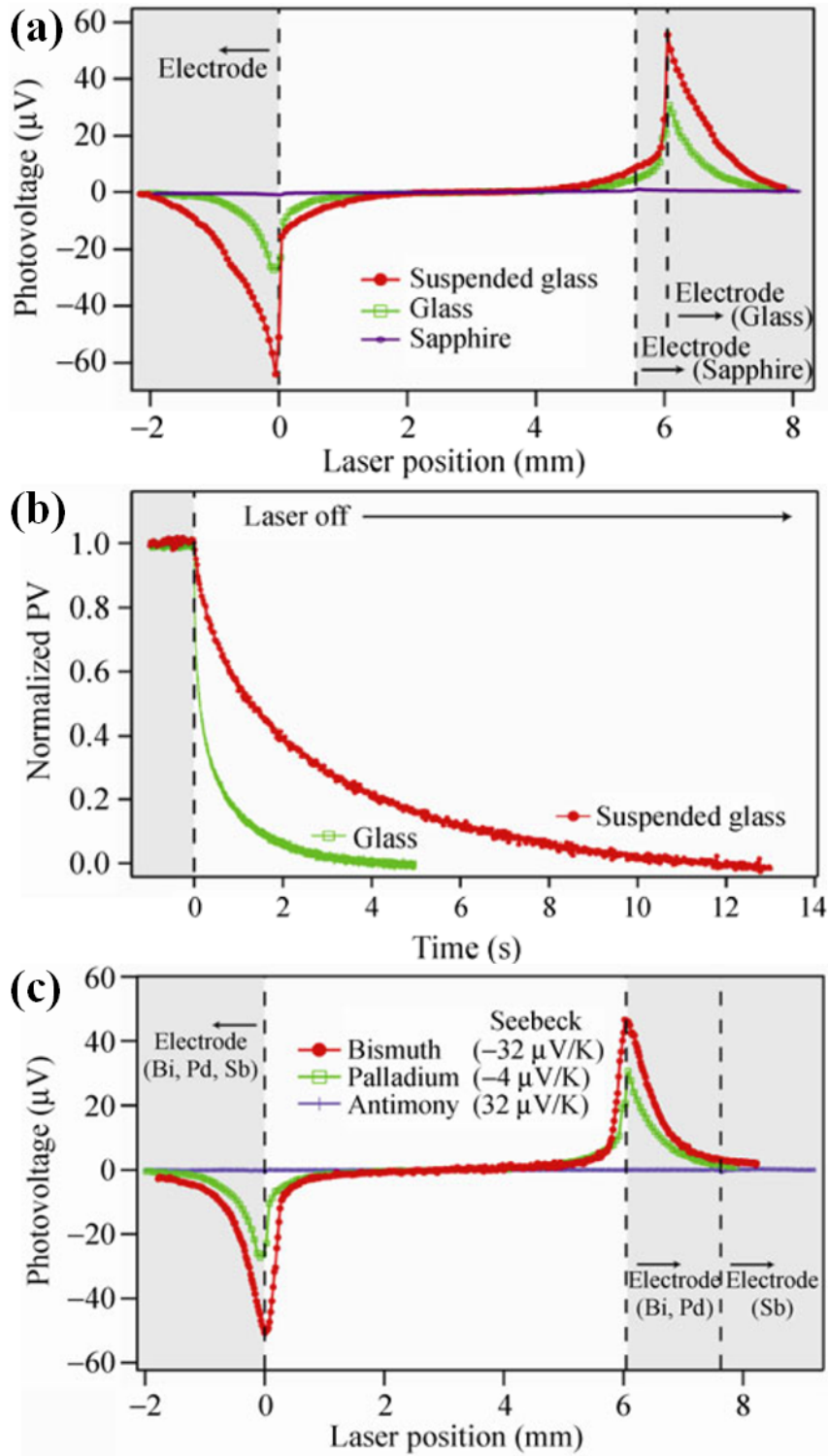

**Figure 14.** a) Substrate dependence of photovoltage on the SWCNT film originated from photothemoelectric (PTE) effect. b) The response time of photovoltage of a SWCNT film deposited on a glass slide, which was suspended (red) or supported (green) on a copper block. c) Dependence of photovoltage of SWCNT films on electrodes made by different materials (bismuth, palladium, and antimony) on glass. Reproduced with permission.[55] Copyright 2011, Springer.

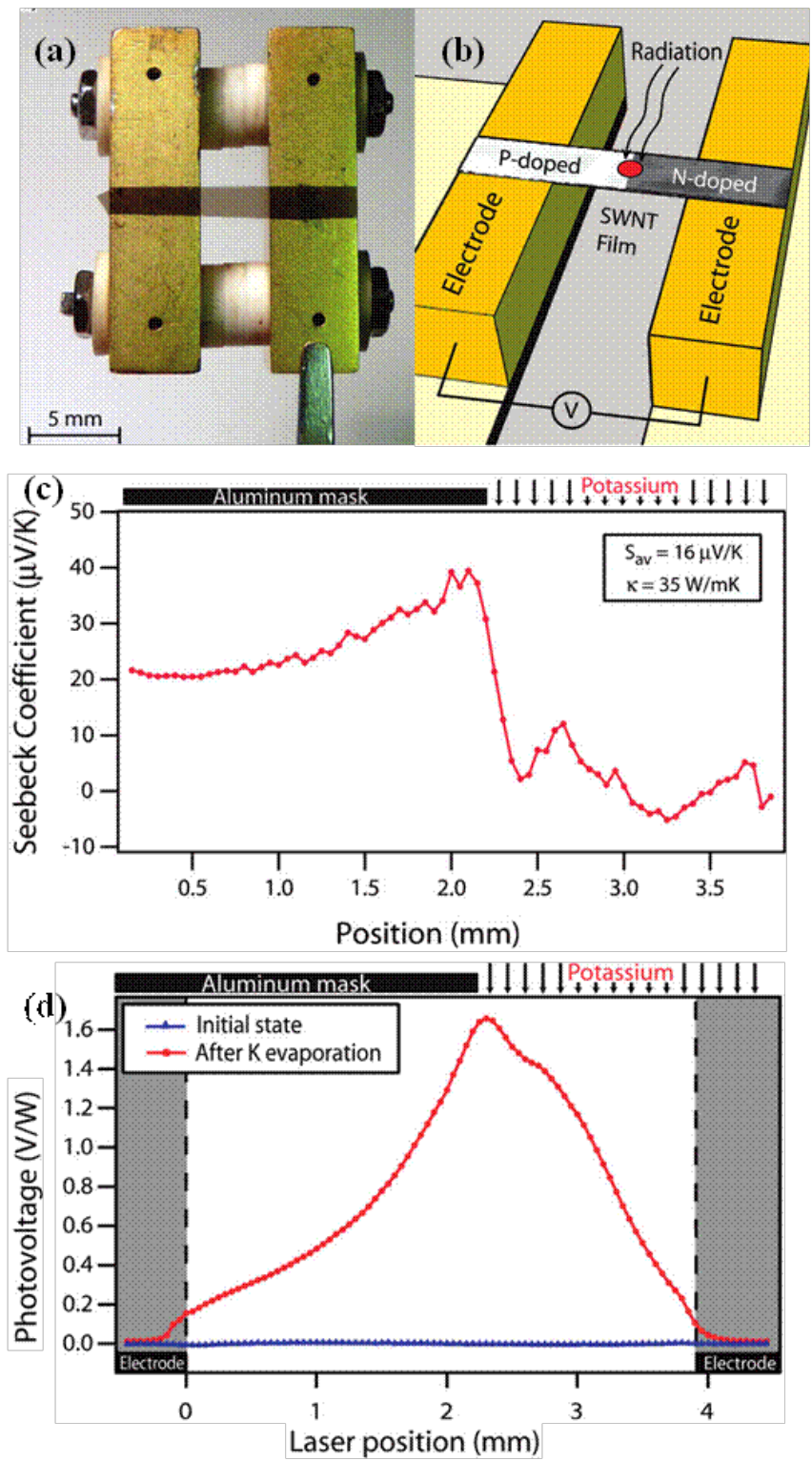

**Figure 15.** a) A suspended SWCNT film of 140 nm thickness bridging a 3.9 mm gap between two gold-plated aluminum electrodes. b) Schematic of SWCNTs thermopile with a p-n doping profile at the center. c) Local Seebeck coefficient profile on the suspended SWCNT film. d) Position dependence of photovoltage response of the thermopile in vacuum. Reproduced with permission.[110] Copyright 2011, American Chemical Society.

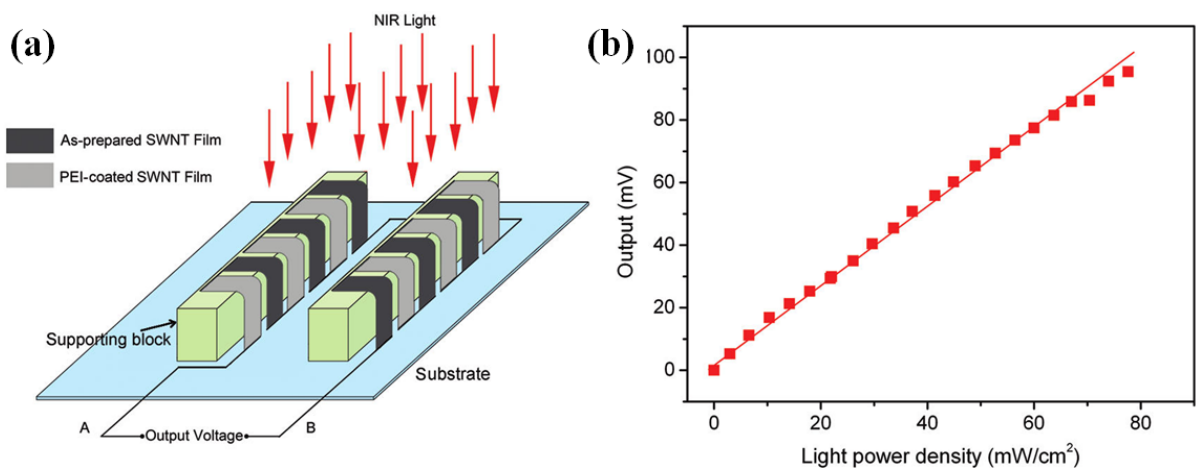

**Figure 16.** a) Schematic image of an IR sensor with multiple p-n segments of SWCNT sheets connected in series. b) IR (980nm) Power density dependence of the photovoltage of the device. Reproduced with permission.[106] Copyright 2010, Aemrican Chemical Society.

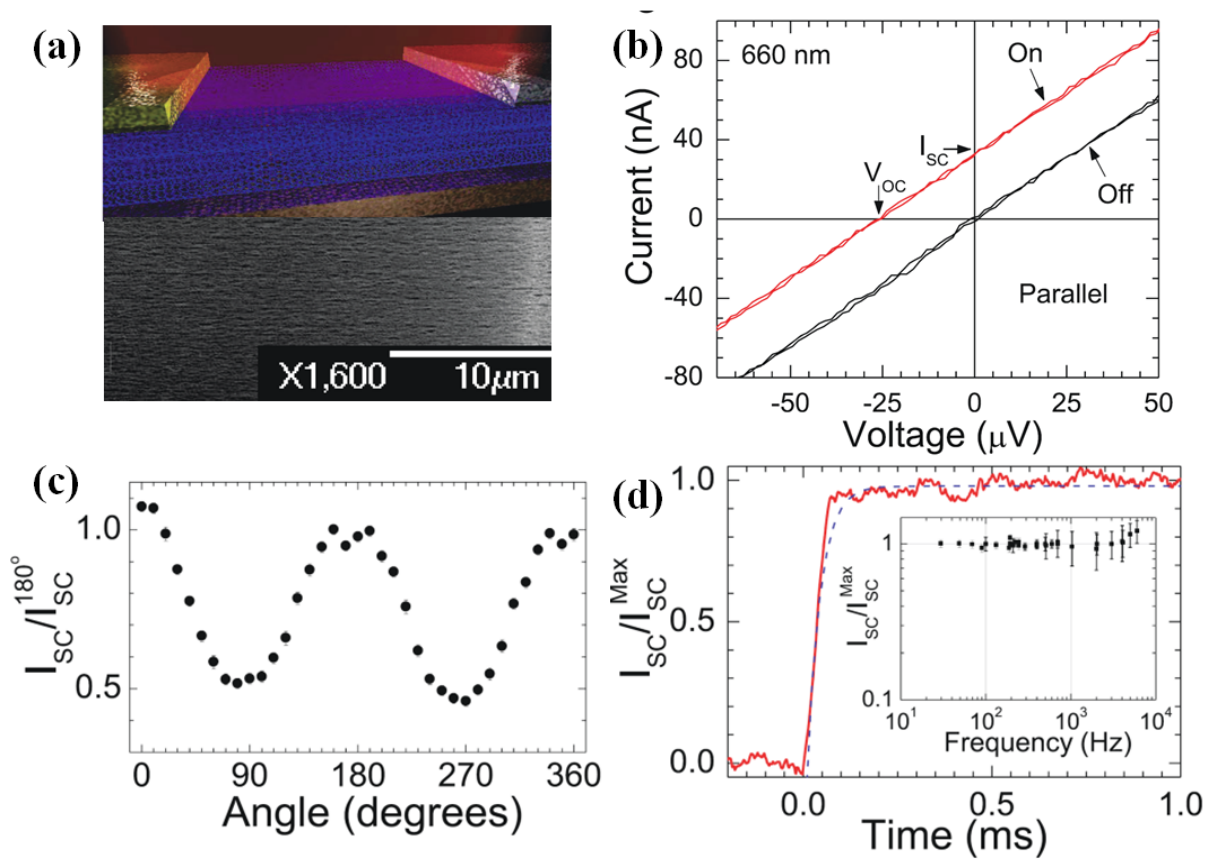

**Figure 17.** a) Schematic of a PTE photodetector based on macroscopically aligned SWCNTS film. Inset show the SEM image of the aligned SWCNT film. b) I-V curves of in the dark and under global illumination at 660 nm on the photodetector with asymmetric contacts. c) Polarization dependence of the photoresponse under local illumination. d) Photocurrent time response under chopped excitation and local illumination. Inset shows the frequency dependence of the photoresponse recorded using a lock-in amplifer. Reproduced with permission.[113] Copyright 2013, Nature Publishing Group.

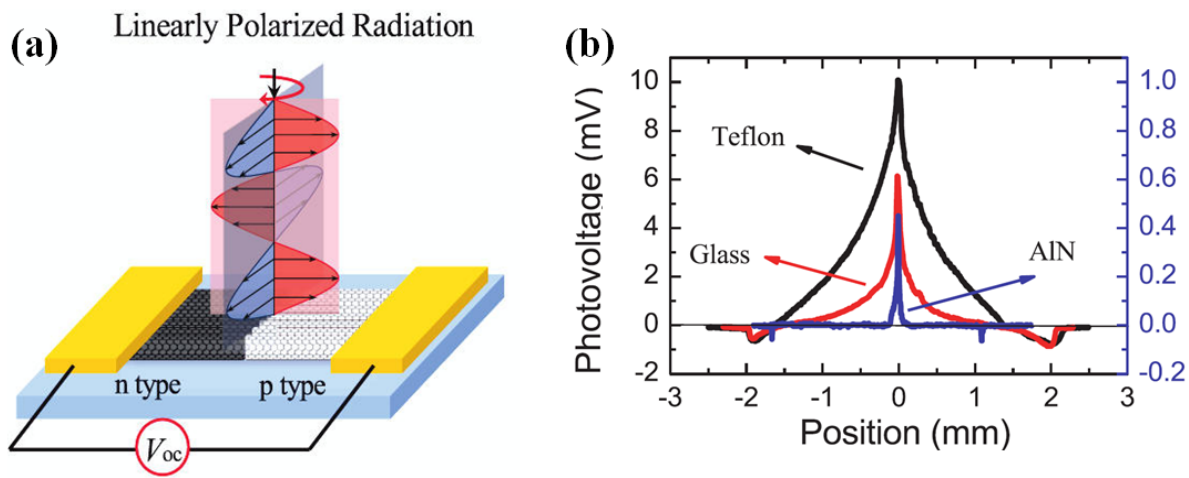

**Figure 18.** a) Schematic of a PTE photodetector based on aligned SWCNTs film with a p-n junction at the center under the illumination of polarized IR light. b) Position dependence of photovoltage on the device with different substrate. Reproduced with permission.[109] Copyright 2013, American Chemical Society.

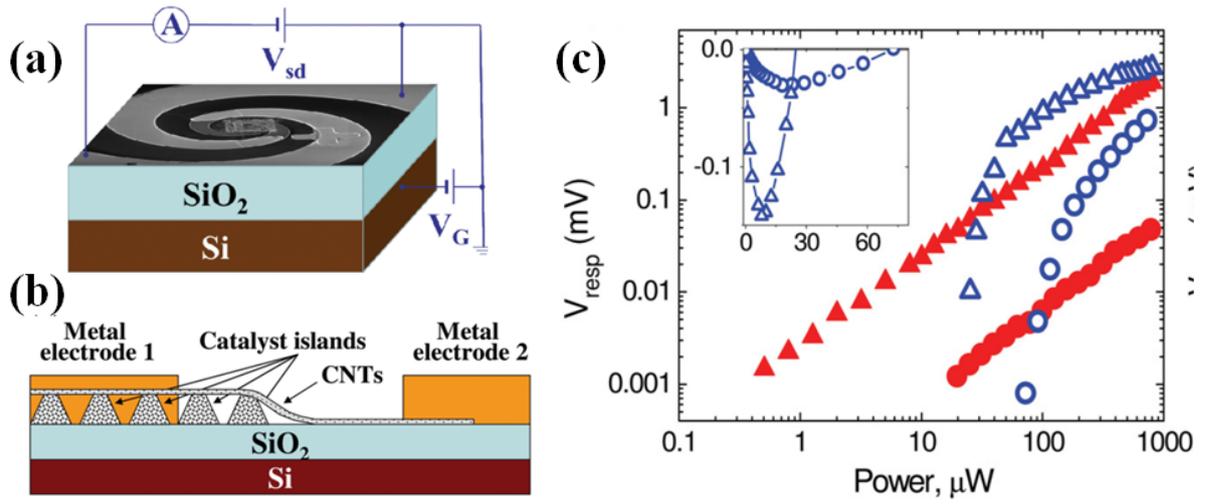

**Figure 19.** a) Schematic of the PTE photodetector based on SWCNTs network with a spiral attenuate structure. b) Schematic image of the SWCNT network with asymmetric thermal contact to the substrates. c) Photovoltage response of two devices at room temperature (filled symbols) and 4.2 K (hollow symbols) as a function the power of incident sub THz radiation (140 GHz). Reproduced with permission.[114] Copyright 2013, AIP Publishing LLC.

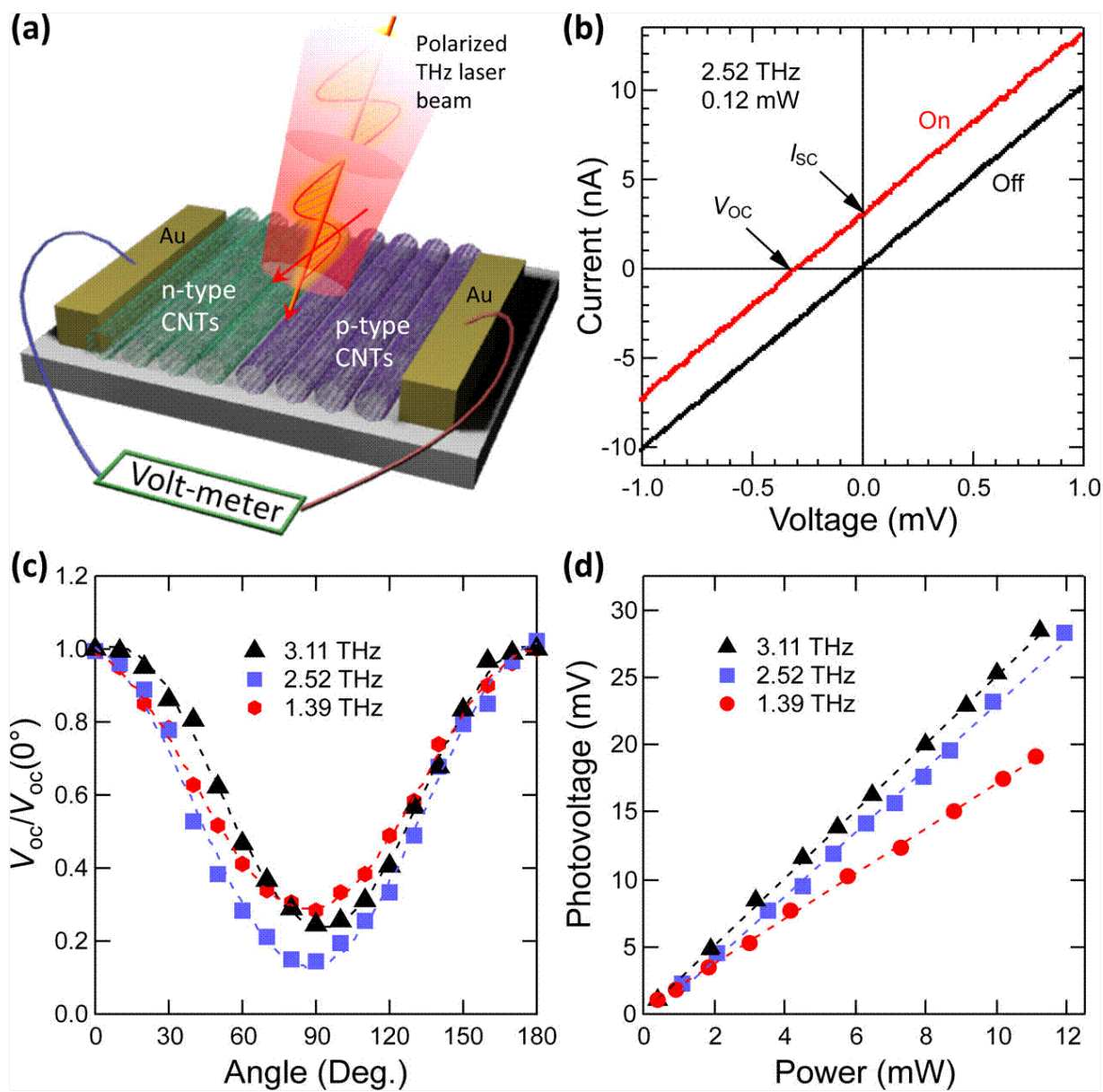

**Figure 20.** Characteristic of a THz detector based on aligned SWCNTs a) Schematic of the photodetector. b) I-V curves of the detector in dark and under the THz illumination with a frequency of 2.52 THz. c) Polarization dependence of Photovoltage at different THz frequency. d) Power dependence of the photovoltage at different THz frequency. Reproduced with permission.[112] Copyright 2014, American Chemical Society.

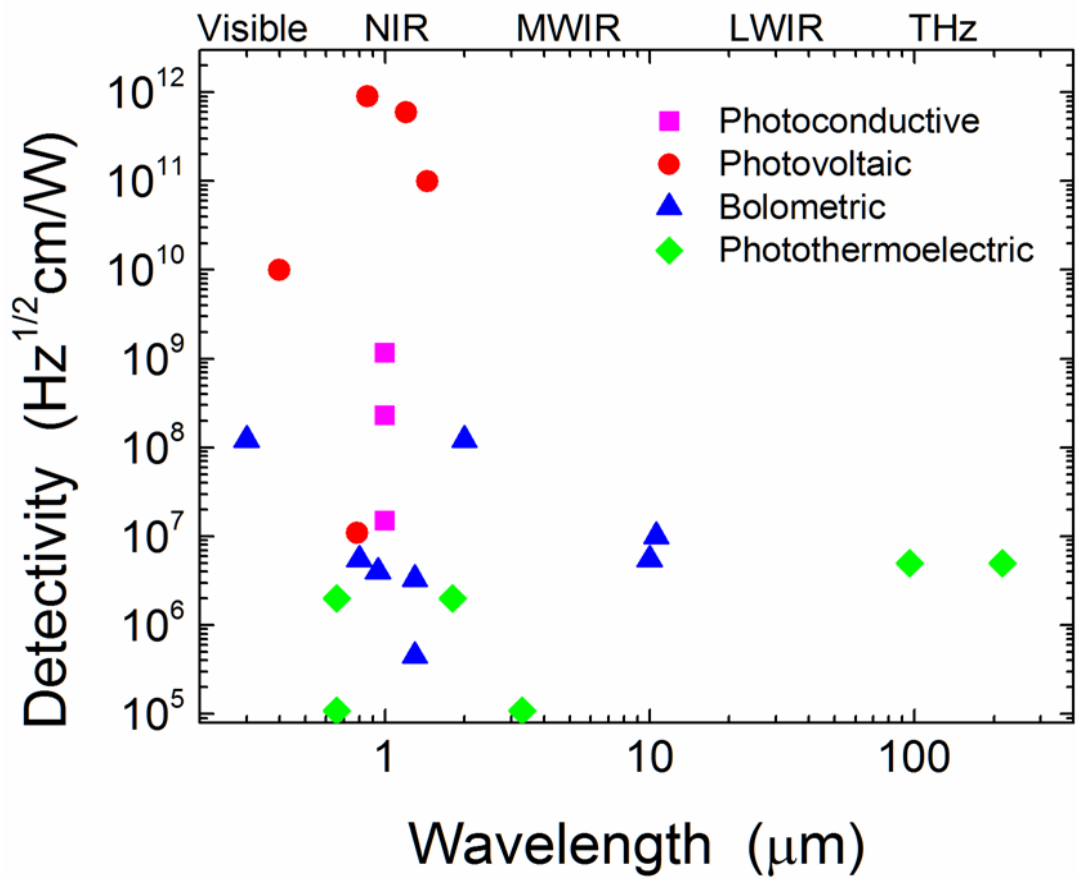

**Figure 21.** Detectivity of various CNT photodetector types in the different regions of the electromagnetic spectrum. The data can be found in the Tables in the text. When the wavelength range in Tables is small, a single data point is indicated in the middle of the range; if the range is larger, two data points are indicated at the ends of the range.